\documentclass{PoS}
\usepackage{graphicx,epsfig,dcolumn,bm,epic,eepic,float}
\usepackage{amsmath}
\usepackage[T1]{fontenc}
\usepackage[utf8]{inputenc}
\usepackage{mathtools} 
\usepackage{pifont}
\usepackage{amsmath}
\usepackage{latexsym}
\usepackage{color}
\usepackage{cancel}
\usepackage{scrextend}
\usepackage{amsmath}
\usepackage{amsthm}
\usepackage{eucal}
\usepackage{amssymb}
\usepackage{mathrsfs}
\usepackage[bottom]{footmisc}
\usepackage{makeidx,shortvrb,latexsym}
\usepackage{epstopdf}
\usepackage{mathtools}
\usepackage{ragged2e}
\usepackage[T1]{fontenc}
\newcommand {\be}{\begin{equation}}
\newcommand {\ee}{\end{equation}}
\newcommand {\ba}{\begin{eqnarray}}
\newcommand {\ea}{\end{eqnarray}}
\newcommand {\bea}{\begin{eqnarray}}
\newcommand {\eea}{\end{eqnarray}}

\pdfoutput=1

\numberwithin{equation}{section}
\title{Natural Alignment in Multi-Higgs Doublet Models}

\ShortTitle{ Natural Alignment in Multi-Higgs Doublet Models}

\author{Neda Darvishi\\
      Consortium for Fundamental Physics, School of Physics and
  Astronomy,\\University of Manchester, Manchester M13 9PL, United
  Kingdom\\
  E-mail: \email{neda.darvishi@manchester.ac.uk}}
\author{\speaker{ Apostolos Pilaftsis}\\
    Consortium for Fundamental Physics, School of Physics and
  Astronomy,\\University of Manchester, Manchester M13 9PL, United
  Kingdom\\
    E-mail: \email{apostolos.pilaftsis@manchester.ac.uk}}

  \abstract{ We present the \textit{complete} set of continuous
    maximal symmetries that the potential of an $n$-Higgs Doublet
    Model ($n$HDM) should satisfy for {\em natural} Standard Model
    (SM) alignment.  As a result, no large mass scales or fine-tuning
    is required for such alignment, which still persists even if these
    symmetries were broken {\em softly} by bilinear mass terms. In
    particular, the Maximal Symmetric $n$HDM (MS-$n$HDM) can provide
    both natural SM alignment and quartic coupling unification up to
    the Planck scale.  Most remarkably, we show that the MS-2HDM is a
    very predictive extension of the SM governed by two only
    additional parameters: (i) the charged Higgs mass $M_{h^{\pm}}$
    (or $m^2_{12}$) and (ii) $\tan\beta$, whilst the quartic coupling
    unification scale~$\mu_X$ is predicted to assume two {\em
      discrete} values. With these two input parameters,
    the entire Higgs-mass spectrum of the model can be determined.
    Moreover, we obtain definite predictions of misalignment for the
    SM-like Higgs-boson couplings to the gauge bosons and to the
    quarks,\,which might be testable at future precision high-energy
    colliders.}

  \FullConference{Corfu Summer Institute 2019 "School and Workshops on Elementary Particle Physics and Gravity" (CORFU2019)\\
    31 August - 25 September 2019\\
    Corfù, Greece}

\begin{document}

\section{Introduction}
\label{sec:intro} 

The quest for new physics beyond the Standard Model (SM), such as the
exploration of non-standard scenarios with extended Higgs sectors, has
strong theoretical and experimental motivations${}$.  The data collected
from the CERN Large Hadron Collider (LHC) impose constraints over the
coupling strengths of the Higgs boson, primarily to the electroweak
(EW) gauge bosons ($Z, \,W^\pm$), which are very close to SM
predictions \cite{ATLAS1,CMS1,Darvishi:2019uzp}. This simple fact restricts severely
the form of possible scalar-sector extensions of the SM.

An interesting class of Higgs-sector extensions is the one that
augments the SM with $n \geq 2$ Higgs doublets, usually called the
$n$-Higgs Doublet Model ($n$HDM).  In the $n$HDM, the couplings of the
SM-like Higgs boson to the EW gauge bosons ($Z,\,W^\pm$) must
resemble those predicted by the SM, so as to be in agreement with the
current Higgs signals at the LHC. This is only possible within the
so-called SM alignment
limit~\cite{Ginzburg:1999fb,Chankowski:2000an,Delgado:2013zfa,Carena:2013ooa,Dev:2014yca,Bernon:2015qea,Darvishi:2017bhf,Benakli:2018vqz,Lane:2018ycs}.

Achieving natural SM alignment via the implementation of symmetries
has a historical background that reaches back as far as the
introduction of the Cabibbo–Kobayashi–Maskawa (CKM)
matrix~\cite{GellMann:1960np,Cabibbo:1963yz,CKM}. If the
three-generation of SM is complete, then the rotation embodied in the
CKM matrix must be unitary.  A similar approach utilises the so-called
Glashow-Iliopoulos-Maiani (GIM)~\cite{Glashow:1970gm} mechanism to
explain the smallness of the strangeness-changing interaction at the
quantum level. Interestingly enough, the GIM mechanism requires the
existence of the $c$-quark, and the conservation of strangeness by
neutral currents naturally follows from the group structure and
representation content of the SM. Later in 1977~\cite{Paschos:1976ay},
the necessary conditions for {\em natural} diagonal neutral currents
in $Z$-boson interactions to all quark fields were proposed, where the
solution was a generalization of the GlM scheme to many quark fields
with two distinct charges. In the same period, equivalent conditions
were derived by the authors in~\cite{Glashow:1976nt}, who extended the
concept of natural diagonal interactions to multi-Higgs-boson
interactions to quarks.  By 1996, it has been shown that the
flavour-changing neutral current (FCNC) couplings of the neutral
scalars of 2HDM can be related to elements of the CKM matrix with the
help of symmetries of the model. From this brief historic review, it
is evident that symmetries were always at the heart for realising {\em
  natural} SM alignment, i.e.~having alignment or good agreement with
SM predictions without decoupling of large mass scales or without
resorting to {\em ad-hoc} arrangements among the parameters of the
candidate new-physics
theory~\cite{Georgi:1978ri,Gunion:2002zf,CPMK,Carena:2013ooa,Haber:2015pua,
  Grzadkowski:2018ohf}.

The potential of $n$HDMs contains a large number of
SU(2)$_L$-preserving accidental symmetries as subgroups of the
symplectic group Sp(2$n)$. The complete set of accidental symmetries
that may occur in the tree-level scalar potential of $n$HDMs is
classified in~\cite{Darvishi:2019dbh}.  In particular, we have identified
all accidental symmetries and derived the relationship among
the theoretical parameters of the scalar potential for: (i) the Two
Higgs Doublet Model (2HDM) and (ii) the Three Higgs Doublet Model
(3HDM). We recover the maximum number of $13$ accidental symmetries
for the 2HDM potential and {\em for the first time}, we presented the
complete list of $40$ accidental symmetries for the 3HDM potential.
Here, we Identify the complete set of continuous maximal symmetries
for SM alignment in potential of $n$HDMs.

As an example, we discuss the phenomenological implications of natural
SM alignment limit for Maximally Symmetric Two-Higgs Doublet Model
(MS-2HDM)~\cite{Lee1,Branco1}.  This minimal model can account for a
SM-like Higgs boson, and contains additionally one charged and two
neutral scalars whose observation could be within reach of the
LHC~\cite{Darvishi:2017bhf,Arbey:2017gmh,Hanson:2018uhf}.  In MS-2HDM,
the aforementioned SM alignment can emerge naturally as a consequence
of a continuous symmetry Sp(4) $\cong$ SO$(5)$ in the Higgs
sector~\cite{Pilaftsis:2011ed,
  Dev:2014yca,Pilaftsis:2016erj,Dev:2017org,Darvishi:2019ltl}. The
SO$(5)$ symmetry can be broken explicitly by two sources: (i)~by
renormalization-group~(RG) effects and (ii)~softly by the bilinear
scalar mass term~$m^2_{12}$. One of the interesting properties of this
model is that all quartic couplings can unify at very large scales
$\mu_X \sim 10^{11}\,$--$\,10^{20}$\,GeV, for a wide range of
$\tan\beta$ values and charged Higgs-boson masses.  Specifically, we
find that quartic coupling unification can emerge in two different
conformally invariant points, where all quartic couplings vanish. The
first conformal point is at relatively low-scale typically of order
$\mu_X^{(1)} \sim 10^{11}$~GeV, while the second one is at high scale
close to the Planck scale $\mu_X^{(2)}\sim 10^{19}$~GeV. Most
remarkably, we show that the MS-2HDM is a very predictive extension of
the SM which is governed by only two additional parameters: (i) the
charged Higgs mass $M_{h^{\pm}}$ (or $m^2_{12}$) and (ii) the ratio
$\tan\beta$ of the two Higgs-doublet vacuum expectation values,
whereas the quartic coupling unification scale~$\mu_X$ takes two
discrete values as mentioned above.  The two parameters, $M_{h^{\pm}}$
and $\tan\beta$, also suffice to determine the entire Higgs-mass
spectrum of the model~\cite{Darvishi:2019ltl}, for given values
of~$\mu^{(1,2)}_X$.  These input parameters enable us to obtain
definite predictions of misalignment for the SM-like Higgs-boson
couplings to the gauge bosons and to the top- and bottom-quarks, which
might be testable at future precision high-energy colliders.

These proceedings are organised as follows. Section \ref{sec:1}
briefly reviews the basic features of the 2HDM and discusses the
conditions for achieving exact SM alignment. In Section \ref{sec:2},
we define the $n$HDMs in the bilinear scalar field formalism. Given
that Sp(2$n)$ is the maximal symmetry of the $n$HDM potential, we
present the complete set of continuous maximal symmetries for SM
alignment that may take place in the $n$HDM potentials.  Then, we
introduce prime invariants to build potentials that are invariant
under SU(2)$_L$-preserving continuous symmetries and present
symmetries.  In Section \ref{sec:3}, we focus on MS-2HDM and outline
the breaking pattern of the SO$(5)$ symmetry, which results from the
soft-breaking mass $m^2_{12}$ and the RG effects. In this section, we
also show that the running quartic couplings can be unified at two
different conformally-invariant points and presents our misalignment
predictions for Higgs-boson couplings to gauge bosons and top- and
bottom-quarks. Finally, Section \ref{con} contains our conclusions.

\section{The 2HDM and SM Alignment} \label{sec:1}

The Higgs sector of the 2HDM is described by two scalar SU(2) doublets, 
\begin{equation}
{\phi}_i\ =\ \left(
 \begin{matrix}
\phi_i^+\\
 \phi_i^0
 \end{matrix}\right),
\end{equation} 
with $i=1,2$. Both doublets have the same U(1)$_Y$-hypercharge quantum number, $Y_{\phi_{ i} }={1/2}$.
In terms of these doublets, the
general SU(2)$_L\times$U(1)$_Y$-invariant Higgs potential is given by
\begin{eqnarray}
  \label{eq:V2HDM}
V &=& - \mu_1^2 ( \phi_1^{\dagger} \phi_1) - \mu_2^2 ( \phi_2^{\dagger} \phi_2) 
 - \Big[ m_{12}^2 ( \phi_1^{\dagger} \phi_2)\: +\: {\rm H.c.}\Big] \nonumber \\
 &+& \lambda_1 ( \phi_1^{\dagger} \phi_1)^2 + \lambda_2 ( \phi_2^{\dagger} \phi_2)^2
 + \lambda_3 ( \phi_1^{\dagger} \phi_1)( \phi_2^{\dagger} \phi_2)
 + \lambda_4 ( \phi_1^{\dagger} \phi_2)( \phi_2^{\dagger} \phi_1) \nonumber \\
 &+& \bigg[\, {1 \over 2} \lambda_5 ( \phi_1^{\dagger} \phi_2)^2
 + \lambda_6 ( \phi_1^{\dagger} \phi_1)( \phi_1^{\dagger} \phi_2)
 + \lambda_7 ( \phi_1^{\dagger} \phi_2)( \phi_2^{\dagger} \phi_2)\: +\:
   {\rm H.c.} \bigg],
\end{eqnarray}
where the mass terms $\mu^2_{1,2}$ and quartic couplings $\lambda_{1,2,3,4}$ are real  parameters.  Instead, the
remaining mass term $m^2_{12}$ and the quartic
couplings $\lambda_{5,6,7}$ are complex. Of these 14 theoretical
parameters, only 11 are physical, since 3 parameters can be removed away
using an SU(2) reparameterisation of the Higgs doublets $\phi_1$ and
$\phi_2$~\cite{CPMK}.
 
 In the case of CP-conserving Type-II 2HDM, both
scalar doublets $\phi_1$ and $\phi_2$ receive real and nonzero vacuum
expectation values (VEVs). In detail, we have
$\langle \phi_1^0\rangle= v_1/\sqrt{2}$ and
$\langle \phi_2^0 \rangle = v_2/\sqrt{2}$, where $t_\beta\equiv\tan\beta =v_2/v_1$ and the VEV of
the SM Higgs doublet is $v \equiv {(v_1^2+v_2^2)}^{1/2}$. 

This model can account for only five physical scalar states: two
CP-even scalars ($h$,$H$), one CP-odd scalar $(a)$ and two charged
bosons ($h^{\pm}$). The masses of the $a$ and $h^\pm$  scalars are given by
\begin{eqnarray}
M_a^{2} \ &=&\ M_{h^{\pm}}^{2} + {v^{2} \over 2} (\lambda_4 - \lambda_5),\nonumber \\
M_{h^{\pm}}^{2} \ &=&\ {m_{12}^{2} \over c_{\beta} s_{\beta} }-{v^{2} \over 2} (\lambda_4 + \lambda_5)
 + {v^{2} \over 2 c_{\beta} s_{\beta} } (\lambda_6 c_{\beta}^{2} 
 + \lambda_7 s_{\beta}^{2}), 
\end{eqnarray}
where $s_{\beta} \equiv \sin{\beta}$, $c_{\beta} \equiv \cos{\beta}$.

The masses of the two CP-even scalars, $h$ and $H$ may be obtained by
diagonalising the $2\times2$ CP-even mass matrix~$M^{2}_S$,
\begin{equation}
M_S^{2}=\left(
\begin{matrix}
{A}\,&\, {C} \\
{C}\,&\, {B}
 \end{matrix}\right),
 \label{abc}
\end{equation}
which may explicitly be written down as
 \begin{eqnarray}
M_S^{2}\ =\ M_a^{2} \left( \begin{matrix}
 s_{\beta}^{2} & - c_{\beta} s_{\beta}  \\
 - c_{\beta} s_{\beta}  & c_{\beta}^{2}
 \end{matrix} \right) 
 + v^{2} \left(
 \begin{matrix}
 2 \lambda_1 c_{\beta}^{2} + \lambda_5 s_{\beta}^{2} + 2 \lambda_6 c_{\beta} s_{\beta}  &
 \lambda_{34} c_{\beta} s_{\beta}  + \lambda_6 c_{\beta}^{2} + \lambda_7 s_{\beta}^{2} \\
 \lambda_{34} c_{\beta} s_{\beta}  + \lambda_6 c_{\beta}^{2} + \lambda_7 s_{\beta}^{2} &
 2 \lambda_2 s_{\beta}^{2} + \lambda_5 c_{\beta}^{2} + 2 \lambda_7 c_{\beta} s_{\beta} 
 \end{matrix}
 \right), \nonumber 
\end{eqnarray}
with $\lambda_{34} \equiv \lambda_{3} + \lambda_{4}$. 
The mixing angle $\alpha$ is required for the diagonalisation of $M_S^{2}$, which may be determined by 
\begin{equation}
 \label{alpha}
\tan 2\alpha\ =\ {2C \over A-B}.
\end{equation}

In the so-called Higgs basis \cite{Georgi:1978ri}, the CP-even mass matrix
$M_S^{2}$ given in~(\ref{abc}) takes on the form
 \begin{eqnarray}
{\widehat{M}_S}^{\,2} & = &  \left(
 \begin{matrix}
 \widehat{A} & \widehat{C} \\
 \widehat{C} & \widehat{B}
 \end{matrix}
 \right) = \left(
 \begin{matrix}
  {c_{\beta}} &  { s_{\beta}} \\
 -{s_{\beta}}  &  {c_{\beta}}
 \end{matrix}
 \right) {M_S}^{2} \left(
 \begin{matrix}
 c_{\beta} & -{s_{\beta}} \\
 s_{\beta} &  c_{\beta}
 \end{matrix}
 \right),
 \label{MS2}
\end{eqnarray}
with
\begin{eqnarray}
\label{Chat}
\widehat{A} & = & 2 v^{2} \left[ c_{\beta}^4 \lambda_1 + s_{\beta}^{2} c_{\beta}^{2} \lambda_{345}
 + s_{\beta}^4 \lambda_2 + 2 c_{\beta} s_{\beta} \left( c_{\beta}^{2} \lambda_6 
 + s_{\beta}^{2} \lambda_7 \right) \right], 
\nonumber \\
\widehat{B} & = & M_a^{2} + \lambda_5 v^{2} + 2 v^{2} \left[ s_{\beta}^{2} c_{\beta}^{2} \left(
 \lambda_1 + \lambda_2 - \lambda_{345} \right) - c_{\beta} s_{\beta}  \left(
 c_{\beta}^{2} - s_{\beta}^{2} \right) \left(\lambda_6-\lambda_7 \right) \right],
 \\ 
\widehat{C} & = & v^{2} \left[ s_{\beta}^3 c_{\beta} \left( 2 \lambda_2 - \lambda_{345} \right)
 - c_{\beta}^3 s_{\beta} \left( 2 \lambda_1 - \lambda_{345} \right)
 + c_{\beta}^{2} \left( 1 - 4 s_{\beta}^{2} \right) \lambda_6
 + s_{\beta}^{2} \left(4 c_{\beta}^{2} - 1 \right) \lambda_7 \right].
\nonumber 
\end{eqnarray}

The SM Higgs field may now be identified by the linear field combination,
\begin{equation}
 \label{g-c}
H_{\text{SM}}\ 
 =\, H \cos (\beta - \alpha) \,+\, h \sin (\beta - \alpha).
\end{equation}
To this extend, one may obtain the SM-normalised couplings of the CP-even scalars, $h$ and $H$,  to the EW gauge bosons ($V = W^{\pm}, Z$) as follows:
\begin{equation}
g_{hVV} = \sin (\beta - \alpha), \qquad g_{HVV} = \cos (\beta - \alpha).
\end{equation}
 In similar way, the SM-normalised couplings of the
CP-even scalars to quarks may be derived. In Table~\ref{tab}, these couplings are displayed. 

\begin{table*}[t]
 \centering
 \small
 \begin{tabular}{l c c c c c c c c c c c }
 \hline\hline
S &&& $g_{SVV}$ ($V=W^{\pm},Z$) &&&& $g_{Suu}$ &&&& $g_{Sdd}$ \\
 \hline
 $h$ &&& $\sin(\beta-\alpha)$ &&&& $ \sin(\beta-\alpha)+t^{-1}_\beta\cos(\beta-\alpha)$ &&&& $\sin(\beta-\alpha)-t_\beta\cos(\beta-\alpha)$ \\
 $H$ &&& $\cos(\beta-\alpha)$&&&&$ \cos(\beta-\alpha)-t^{-1}_\beta\sin(\beta-\alpha)$ &&&& $\cos(\beta-\alpha)+t_\beta\sin(\beta-\alpha)$\\
 \hline\hline
 \end{tabular}
 \caption{\it
 Tree-level couplings of CP-even scalar boson $S$ (with$\,S = h,\,
 H$) to the EW gauge bosons ($Z$,\,$W^\pm$)
 and to up-type and down-type quarks in the Type-II~2HDM.}\label{tab}
\end{table*}

From~Table~\ref{tab}, we observe that there are two scenarios to realise
the SM alignment limit:
\begin{itemize}
\item[$\bullet$] SM-like ~$H$ scenario:
$M_H \approx 125\text{~GeV},\,
\cos(\beta-\alpha) \to 1,\,\,  \text{with} \,\, \beta\approx\alpha$.
\item[$\bullet$] SM-like ~$h$ scenario: 
$M_h \approx 125\text{~GeV},\,  \sin(\beta-\alpha) \to 1 ,\,\,  \text{with} \,\,  \beta-\alpha \approx \pi/2$.
\end{itemize}
In these limits, the CP-even $H\, (h)$ scalar couples to the EW gauge
bosons with coupling strength exactly as that of the SM Higgs boson,
while $h\, (H)$ does not couple to them at all~\cite{Dev:2014yca}.  In
the above two scenarios, the SM-like Higgs boson mass is identified with
the $\sim 125~\text{GeV}$ resonance observed at the LHC
\cite{Aad:2012tfa,Chatrchyan:2012xdj}.  In the literature, the neutral
Higgs partner $(H)$ in the SM-like~$h$ scenario is usually termed the
heavy Higgs boson. Instead, in the SM-like~$H$ scenario, the partner
particle $h$ can only have a mass smaller than $\sim\,125~\text{GeV}$
~\cite{Bernon:2015qea}.  In our study, we adopt the SM-like~$H$
scenario, but the partner $h$ would be either heavier or lighter than
the observed scalar resonance at the LHC.

In~\eqref{Chat}, the SM alignment limit $\cos(\beta-\alpha)~\to~1$ can be achieved in
two different approaches: (i)~$\widehat{C} \to 0$ and
(ii)~$M_{h^\pm}\!\sim\!M_a \gg v$.  The first realisation~(i) does not
depend on the choice of the non-SM scalar masses, such as $M_{h^\pm}$
and $M_a$, whereas the second one~(ii) is only possible in the
well-known decoupling
limit~\cite{Georgi:1978ri,Gunion:2002zf,CPMK}.  In
the first realisation, SM alignment is obtained by setting $\widehat{C}=0$,
which in turn implies the condition~\cite{Dev:2014yca}:
 \begin{align}
  \label{A-C}
\lambda_7 t_{\beta}^4 - \left( 2 \lambda_2 - \lambda_{345} \right) t_{\beta}^3 
+ 3 \left( \lambda_6 - \lambda_7 \right) t_{\beta}^{2} 
+ \left( 2 \lambda_1 - \lambda_{345} \right) t_{\beta} - \lambda_6\ =\ 0.
\end{align}
Barring fine-tuning among quartic couplings, \eqref{A-C} leads to
the two different types of constraints:
\begin{enumerate}
\item{
\begin{align}
  \label{eq:NAcond}
\lambda_1\ =\ \lambda_2\ =\ \frac{\lambda_{345}}{ 2}, \qquad 
\lambda_6\ =\ \lambda_7\ =\ 0,
\end{align}
which are independent of $\tan\beta$ and non-standard scalar masses.
In this case, the masses of both CP-even scalars in the alignment limit are given by
\begin{align}
M_H^{2}\ & = \ 2 v^{2} \left( \lambda_1 c_{\beta}^4 + \lambda_{345} s_{\beta}^{2} c_{\beta}^{2}
 + \lambda_2 s_{\beta}^4 \right)\ \equiv\ 2 \lambda_{\text{SM}} v^{2}, \\ 
M_h^{2}\ & = \ M_a^{2} +  v^{2} \lambda_5 + 2v^{2} s_{\beta}^{2} c_{\beta}^{2} \left(
 \lambda_1 + \lambda_2 - \lambda_{345} \right).
\end{align}}
\item{
\begin{align}
  \label{eq:NAcond2}
\tan\beta=1\:,\quad \lambda_1\ =\ \lambda_2, \quad \ \lambda_{3}, \quad \ \lambda_{4}, \quad \ \lambda_{5}, \quad 
\lambda_6\ =\ \lambda_7\ ,
\end{align}
where the masses of both CP-even scalars in the alignment limit may take the following forms,
\begin{align}
M_H^{2}\ & = \ 2 v^{2} \left( \lambda_1 +  \lambda_2 +  \lambda_{345} + 2 (  \lambda_6 
 +  \lambda_7 )\right)\ , \\
M_h^{2}\ & = \ M_a^{2} + \lambda_5 v^{2} + 2 v^{2} \left(
 \lambda_1 + \lambda_2 - \lambda_{345} \right)\ .
\end{align}} 
\end{enumerate}
Note that the constraints given in 2 represent a
particular limit of the constraints 1. As we will discuss further in
Section~\ref{sec:2}, these constraints mainly lead to an inert Type-I 2HDM in the
Higgs basis, for which SM alignment is automatic thanks to an unbroken
$Z_2$ symmetry.

In the second realisation for the SM alignment limit
$\cos(\beta-\alpha) \to 1$ mentioned above,
i.e. ~$M_{h^\pm}\!\sim\!M_a \gg v$, we may simplify matters by
expanding $M^{2}_{H,h}$ in powers of $v/M_a\ll 1$. So, we may find~\cite{Dev:2014yca}
\begin{align}
M_H^{2}\ &\simeq\ 2\lambda_{\text{SM}} v^{2} - {v^4 s_{\beta}^{2} c_{\beta}^{2}
 \over M_a^{2} + \lambda_5 v^{2}} \,
 \Big[ s_{\beta}^{2} \left( 2 \lambda_2 - \lambda_{345} \right)
 - c_{\beta}^{2} \left( 2 \lambda_1 - \lambda_{345} \right) \Big]^{2}\,,\\
M_h^{2}\ &\simeq\ M_a^{2} + \lambda_5 v^{2}\ \gg\ v^{2}.
\end{align}
Note that for large values of $\tan\beta$, the phenomenological properties of the
$H$-boson become more and more close to those of the SM Higgs
boson~\cite{Carena:2013ooa,Delgado:2013zfa}. Since we are interested in
analysing the misalignment of the $H$-boson couplings from their SM
values, we follow an approximate approach inspired by the seesaw
mechanism~\cite{Minkowski:1977sc}. Particularly, we will express all
the $H$-boson couplings in terms of the light-to-heavy~scalar-mixing
parameter $\theta_\mathcal{S}\equiv\widehat{C}/\widehat{B}$.  Thus, operating~\eqref{alpha}
for the hatted quantities and employing ${\widehat{A}}\gg{\widehat{C}}$,  the following approximate analytic expressions may be derived
\begin{subequations} \label{ex-gc}
\begin{align}
g_{HVV}&\simeq 1-{\theta_\mathcal{S}^{2}\over 2},
\\
g_{hVV}&\simeq -{\theta_\mathcal{S}}\ =\ 
{v^{2} c_{\beta} s_{\beta}  \over M_a^{2} +  v^{2} \lambda_5 }\,
 \Big[\, c_{\beta}^{2} \left( 2 \lambda_1 - \lambda_{345} \right) -
   s_{\beta}^{2} \left( 2 \lambda_2 - \lambda_{345} \right)
   \Big]. 
\end{align}
\end{subequations}
Given the narrow experimental limits on the deviation of $g_{HVV}$
from~1, one must have the parameter
$\theta_\mathcal{S} \ll 1$, which justifies our seesaw-inspired
approximation. In fact, the mixing parameter $\theta_\mathcal{S}$
vanishes in the exact SM alignment limit as
$\alpha \to \beta$.  

To this extent, we derive approximate analytic expressions for the 
$h$- and $H$-boson couplings to up- and down-type quarks.
To leading order in the light-to-heavy scalar-mixing~$\theta_\mathcal{S}$, these are given by
\begin{align}
g_{Huu}&\simeq  1+{t^{-1}_\beta}\,{\theta_\mathcal{S}}, \qquad \qquad
 g_{Hdd} \simeq 1-{\theta_\mathcal{S}}\,{t_\beta}, \\ \nonumber
 g_{huu}&\simeq -{\theta_\mathcal{S}}+{t^{-1}_\beta}, \qquad \qquad
 g_{hdd} \simeq -{\theta_\mathcal{S}}-{t_\beta} .
   \label{mfc}
\end{align}
In the SM alignment limit, we have $g_{Huu}\,\to\,1$ 1 and
$g_{Hdd} \to$ 1.  Obviously, any deviation of these couplings from their SM values is governed by quantities $\tan\beta$ and $\theta_\mathcal{S}$.

In this study, our primary interest lies in natural
realisations of SM alignment, for which neither a mass hierarchy
$M_{h^\pm}\!\sim\!M_a \gg v$, nor a fine-tuning among the quartic couplings would be
necessary.  To this end, one is therefore compelled to identify
possible maximal symmetries of the $2$HDM potential that would impose
the condition stated in~\eqref{A-C}. Thereafter, this result may be generalized for $n$HDM potential
to achieve the SM alignment limit. In~the next section, we will show how SM alignment can be
achieved naturally by virtue of accidental continuous symmetries imposed on the theory.

\section{Multi-Higgs Doublet Models and Natural Alignment} \label{sec:2}

The $n$HDMs contains $n$ scalar doublet fields,
$ \phi_{ i} \,\, (i=1,2,\cdots,n)$, which all have the same
U(1)$_Y$-hypercharge quantum number.  The most general
SU(2)$_L\times$U(1)$_Y$ invariant $n$HDM potential may conventionally
be given as \cite{Botella:1994cs}:
\begin{equation}
V_n=\ \displaystyle\sum_{i,j=1}^{n}\, m_{ij}^{2}\, ( \phi_i^{\dagger} \phi_j)+ \ \displaystyle\sum_{i,j\, k,l=1}^{n}\, 
\lambda_{ijkl}\, ( \phi_i^{\dagger} \phi_j)( \phi_k^{\dagger} \phi_l),
\end{equation}
with $\lambda_{ijkl}=\lambda_{klij}$.
In general, the above potential contains $n^{2}$ physical mass terms and $n^{2}(n^{2}+1)/2$ physical quartic couplings. 

An equivalent way to write the $n$HDM potential is based on the
so-called bilinear field formalism
\cite{Maniatis:2006fs,Nishi:2006tg,Ivanov:2006yq,Battye:2011jj}.  To
this end, we first define a $4n$-dimensional ($4n$-D) complex
${\bm{\Phi}}_n$-multiplet as
\begin{align}
\label{eq:bfPhi}
&{\bm{\Phi}}_2^{\mathsf{T}}  =\begin{pmatrix}
\phi_1, \,
\phi_2, \, 
\tilde{\phi_1}, \,
\tilde{\phi_2} \,
\end{pmatrix}^{\mathsf{T}},
 \nonumber \\
&{\bm{\Phi}}_3^{\mathsf{T}}  =\begin{pmatrix}
\phi_1, \,
\phi_2, \,
\phi_3, \,
\tilde{\phi_1}, \,
\tilde{\phi_2},  \,
\tilde{\phi_3} \,
\end{pmatrix}^{\mathsf{T}} ,
 \nonumber \\
&\cdots
\nonumber \\
&{\bm{\Phi}}_n^{\mathsf{T}}  =\begin{pmatrix}
\phi_1, \,
\phi_2, \,
\phi_3, \,
\cdots, 
\tilde{\phi_1}, \,
\tilde{\phi_2},  \,
\tilde{\phi_3},  \,
\cdots
\end{pmatrix}^{\mathsf{T}} ,
\end{align}
where $\sigma^{1,2,3}$ are the Pauli matrices and $\tilde{\phi_i}=i \sigma^{2} \phi_i^{*}$ are the U(1)$_Y$ hypercharge-conjugate of $\phi_i$.
Observe that the ${\bm{\Phi}}_n$-multiplet transforms covariantly under an SU(2$)_L$ gauge transformation as,
\begin{equation}
{\bm{\Phi}}_n\to \text{U}_L {\bm \Phi}_n, \quad \text{U}_L\in \text{SU(2)}_L.
\label{su(2)}
\end{equation}
Additionally, this multiplet  satisfies the following Majorana-type property \cite{Battye:2011jj},
\begin{align}
{\bm{\Phi}}_n=C {{\bm \Phi}_n^*}, 
\end{align}
where $C=\sigma^{2} \otimes {\bf{1}}_n \otimes \sigma^{2}$ ($C=C^{-1}=C^*$) is the charge conjugation operator and ${\bf{1}}_n$ is the $n\times n$ identity matrix.

With the help of the ${\bm{\Phi}}_n$-multiplet, we may now define the bilinear fields vector~\cite{Ivanov:2006yq,Pilaftsis:2011ed,Battye:2011jj},
\begin{equation}
R_{n}^{A} \equiv {\bm{\Phi}}_n^{\dagger} {\Sigma}^{A}_n {\bm{\Phi}}_n,
\end{equation}
with ${A}=0,1,2,\cdots,n(2n-1)-1$. Notice that $n(2n-1)$-vector $R_{n}^{A}$ is invariant under SU(2)$_L$ transformations thanks to (\ref{su(2)}).

The ${\Sigma}^{A}_n$ matrices have $4n \times 4n$ elements and
can be expressed in terms of double tensor products as
\begin{align}
{\Sigma}^{A}_n=\left(\sigma^0 \otimes t_S^a \otimes \sigma^0,
 \quad \sigma^i \otimes t_A^b \otimes \sigma^0 \right),
\end{align}
where $t_S^a$ and $ t_A^b$ are the symmetric and anti-symmetric matrices of the
SU($n$) symmetry generators, respectively. 

With the aid of $n(2n-1)$-vector $R_{n}^{A}$, the potential $V_n$ for
an $n$HDM can be written down in the quadratic form as
\begin{equation}
V_n=-\dfrac{1}{2}M^n_{A} R_n^{A}+\dfrac{1}{4}L^n_{{A}{A'}}R_n^{A}R_n^{{A'}},
\label{VB}
\end{equation}
where $M^n_{A} $ is the $1 \times n(2n-1)$-dimensional mass matrix and
$L^n_{{A}{A'}}$ is a quartic coupling matrix with
$n(2n-1) \times n(2n-1)$ entries.  Evidently, for a U(1$)_Y$-invariant
$n$HDM potential the first $n^{2}$ elements of $M^n_{A} $ and
$n^{2} \times n^{2}$ elements of $L^n_{{A}{A'}}$ are only relevant, since
the other U(1$)_Y$-violating components vanish.

The gauge-kinetic term $T_n$ is given by
\begin{align}
T_n=\frac{1}{2}(D_{\mu}{\bm{\Phi}}_n)^{\dagger} (D^{\mu}{\bm{\Phi}}_n),
\end{align}
where the covariant derivative, $D_{\mu}$, in the ${\bm{\Phi}}_n$ space is
\begin{align}
 D_{\mu}&= \sigma^0 \otimes {\bf{1}}_n \otimes
          \sigma^0\,\partial_{\mu} \: +\: i\frac{g_w}{2} \sigma^0 \otimes {\bf{1}}_n
          \otimes \sigma^i\,W_{\mu}^i\: +\:  
i{g_Y\over 2}\,B_{\mu}\, \sigma^3 \otimes {\bf{1}}_n \otimes \sigma^0\; .
\end{align}
In the limit $g_Y\to 0$, the gauge-kinetic term is invariant under
$\text{Sp}(2n)/Z_2 \otimes \text{SU}(2)_L$ transformations of the
multiplet-$\bm{\Phi}_n$. In general, the maximal symmetry group acting
on the ${\bm{\Phi}}_n$-space in the $n$HDM potentials is
\begin{equation*}
\text{G}_{\mathsf{n\text{-HDM}}}^{{\bm{\Phi}}_n}=\text{Sp}(2n)/Z_2 \otimes \text{SU}(2)_L,
\end{equation*}
which leaves the local SU($2)_L$ gauge kinetic term of ${\bm{\Phi}}_n$
canonical.  The local $SU(2)_L$ group generators can be represented as
$\sigma^0 \otimes {\bf{1}}_n \otimes (\sigma^{1,2,3}/2)$, that commute
with all generators of Sp($2n)$.

Knowing that Sp(2$n)$ is the maximal symmetry group allows us to
classify all SU(2)$_L$-preserving accidental symmetries of $n$HDM
potentials.  The potential of $n$HDMs contains a large number of
SU(2)$_L$-preserving accidental symmetries as subgroups of the
symplectic group Sp(2$n)$. The complete set of accidental symmetries
that may occur in the tree-level scalar potential of $n$HDMs is
classified in~\cite{Darvishi:2019dbh}. In the same context, we
identify the complete set of continuous maximal symmetries that an
$n$HDM potential must obey for having SM alignment. First, we will
presnt all possible maximal SM-alignment symmetries for the $2$HDM
potential, and then we will generalize these for $n$HDM potentials.

In Section~\ref{sec:1}, it has been shown that the SM alignment in
the $2$HDM potential can be achieved when the constraint~\eqref{A-C}
is fulfilled. This constraint may be due to some accidental
continuous symmetry imposed on the model. Given all accidental
continuous symmetries~\cite{Pilaftsis:2011ed,Darvishi:2019dbh}, we may
look for those symmetries that would impose the condition stated
in~\eqref{A-C}.  Along these lines, the following three symmetries for
SM alignment have been identified satisfying the
condition~\eqref{eq:NAcond}:
\begin{subequations} 
\label{AS1}
\begin{eqnarray}
\text{(i)}~&& \text{SO(2)}:  \quad \mu_1^{2}=\mu_2^{2}; \quad   m_{12}^{2}=0; \quad   \lambda_1\ =\ \lambda_2\ =\ \lambda_{345}/2\,; \quad \lambda_{6,7}\ =\ 0, \\
\text{(ii)}~& &\text{SU(2)}:  \quad \mu_1^{2}=\mu_2^{2}; \quad   m_{12}^{2}=0; \quad \lambda_1\ =\ \lambda_2\ =\ \lambda_{34}/2; \quad 
\ \lambda_{5,6,7}\ =\ 0,\\
\text{(iii)}~&& \text{Sp(4)}: \quad \mu_1^{2}=\mu_2^{2}; \quad   m_{12}^{2}=0; \quad  \lambda_1\ =\ \lambda_2\ =\ \lambda_{3}/2; \quad  \lambda_{4,5,6,7}\ =\ 0.
\end{eqnarray}
\end{subequations} 
In the above list, having SM alignment limit is independent of values
of $\tan\beta$, and the values of the soft-breaking bilinear mass
terms $\mu^{2}_{1,2}$ and $m^{2}_{12}$. In addition to above
symmetries, in the weak basis $\lambda_6 = \lambda_7$,
the possibility $\tan\beta=1$ gives rise to the following
symmetries for SM alignment~\eqref{eq:NAcond2}:
\begin{subequations} 
\label{AS2}
\begin{eqnarray}
\text{(a)}~& &\text{CP2}:   \, \mu_1^{2}=\mu_2^{2}; \, m_{12}^{2}=0; \, \lambda_1=\lambda_2; \,\lambda_3; \, \lambda_4; \,\lambda_5; \,\lambda_{6,7} =0,\\
\text{(b)}~&& (\text{CP1} \rtimes \,S_2)\otimes\text{ Sp(2)}_{\phi_1+\phi_2}:   \, \mu_1^{2}=\mu_2^{2}; \, m_{12}^{2}; \, \lambda_1=\lambda_2; \,\lambda_3; \,\lambda_4=\lambda_5; \,\lambda_6=\lambda_7 ,\\
\text{(c)}~ && (S_2\rtimes Z_2)\otimes \text{Sp(2)}_{\phi_1+\phi_2} :   \, \mu_1^{2}=\mu_2^{2}; \, m_{12}^{2}=0; \, \lambda_1=\lambda_2; \,\lambda_3; \,\lambda_5=\pm \lambda_4;  \,\lambda_{6,7}=0,\\
\text{(d)}~ & &\text{U(1)}_{\text{}}\otimes \text{Sp(2)}_{\phi_1\phi_2} :   \, \mu_1^{2}=\mu_2^{2}; \, m_{12}^{2}=0; \, \lambda_1=\lambda_2=\lambda_3/2; \,\lambda_4; \,  \lambda_{5,6,7}\ =\ 0,\\
\text{(e)}~ && S_2\,\otimes \text{Sp(2)}_{\phi_1}\otimes \text{Sp(2)}_{\phi_2} :   \, \mu_1^{2}=\mu_2^{2}; \, m_{12}^{2}=0; \, \lambda_1=\lambda_2; \,\lambda_3; \;  \,\lambda_{4,5,6,7}=0,
\end{eqnarray}
\end{subequations} 
where the subscript ${\phi_1+\phi_2}$ shows an Sp(2) transformation
that acts on both $(\phi_1, i \sigma^{2} \phi_1^* )^{\mathsf{T}}$ and
$ ( \phi_2, i \sigma^{2} \phi_2^* )^{\mathsf{T}}$. Additionally, the
subscript ${\phi_1 \phi_2}$ denotes an Sp(2) transformation acting on
$( \phi_1, i \sigma^{2} \phi_2^* )^{\mathsf{T}}$. 

It is important to note here that SM alignment does not get spoiled
when $m^{2}_{12} \neq 0$ for the symmetries~(a)--(e) stated
in~\eqref{AS2}. Nevertheless, one should always have $\mu^{2}_1 = \mu^{2}_2$ at the
tree level in order to satisfy the minimization conditions for the
2HDM potential~\cite{Pilaftsis:1999qt}. In the so-called Higgs basis,
all symmetries in~\eqref{AS2}, with exception the one in~(b), lead to restricted forms
of inert Type-I 2HDM potentials, for which SM alignment is automatic
as a result of an unbroken $Z_2$ symmetry.

To sum up, the set of symmetries in ~\eqref{AS1} satisfy alignment conditions
naturally without imposing any constraint on the values of
$\tan\beta$, nor on the bilinear mass terms $\mu^{2}_{1,2}$ and
$m^{2}_{12}$. However, in the set of symmetries ~\eqref{AS2}, exact alignment
can be achieved only for the specific value $\tan\beta=1$ and for
$\mu^{2}_1 = \mu^{2}_2$. 

Having obtained these symmetries, it is straightforward to generalize
these results to the $n$HDM potentials, with $n > 2$. Of the $n$
scalar doublets, we assume that a number $m < n$ correspond to inert
doublets, which do not participate in electroweak symmetry breaking (EWSB),
need to be treated differently.

The full scalar potential $V$ of a naturally aligned $n$HDM may be
written down as a sum of three terms:
 \bea V=V_{sym}+V_{inert}+\Delta
V.
\label{sym+inert}
\eea

We first construct the symmetry-constrained part of the scalar
potential $V_{sym}$ in terms of fundamental building blocks that
respect the symmetries.  Here, we introduce the invariants $S_n$,
$D_n^{2}$ and $T_n^{2}$. In detail, $S_n$ is defined as
\begin{align}
 \label{eq:spn}
 S_n\ =\ {\bm{\Phi}}_n^\dagger{\bm{\Phi}}_n,
\end{align}
which is invariant under both the SU($n)_L\otimes$U(1)$_Y$ gauge group and Sp(2$n$).
Moreover, we define the SU(2)$_L$-covariant quantity $D^a_n$ in the HF 
space as
\begin{equation}
  \label{su}
D_n^a\ =\ {{\Phi}}^\dagger \sigma^a {{\Phi}},
\end{equation}
with $\Phi=(\phi_1,\phi_2,\dots,\phi_n)^{\mathsf{T}}$.
Under an SU(2)$_L$ gauge transformation,
$D^a_n \to D_n^{\prime a}= O^{ab}D_n^b$, where $O\in$~SO(3).  Hence,
the quadratic quantity $D_n^{2}\equiv D^a_n D^a_n$ is both gauge and SU($n$)
invariant. Finally, we define the auxiliary quantity $T_n$ in the HF
space as
\begin{equation}
T_n\ =\ \Phi \Phi^{\mathsf{T}},
\end{equation}
which transforms as a triplet under SU(2)$_L$,
i.e.~$T_n \,\to\, T_n^\prime=U_L T_n U_L^{\mathsf{T}}$.  As a consequence, a
proper prime invariant may be defined as
$\,T_n^{2}\equiv \text{Tr}(TT^*)$, which is also both gauge and SO($n$)
invariant.

Thus, we can construct the symmetry-constrained part of the scalar
potential $V_{sym}$ in terms of prime invariants in the following
form,
\begin{align}
V_{sym}=-\mu ^{2} S_n +\lambda_S S_n^{2}+\lambda_D D_n^{2}+ \lambda_T T_n^{2}.
\end{align}
Obviously, the simplest form of the $n$HDM potentials belong to the
maximal symmetry Sp(2$n$), which has the same form as the SM
potential,
$$
V_{\text{SM}}=-\mu^{2} \left( \phi^{\dagger}\phi \right) 
 + \lambda \left( \phi^{\dagger}\phi\right)^{2},
$$
with a single mass term and a single quartic coupling. For example, the 2HDM Sp(4)${/Z_2}$-invariant potential, the so called MS-2HDM is
\begin{equation}
V_{\text{MS-2HDM}}\ =\ - \mu_1^{2} \left( \left| \phi_1 \right|^{2}\: +\: \left| \phi_2 \right|^{2} \right) 
 + \lambda_1 \left( \left| \phi_1 \right|^{2} + \left| \phi_2 \right|^{2}
  \right)^{2},
  \label{eq:VMS2HDM}
\end{equation}
where the parameters have the following relations,
\begin{align}
\mu_1^{2} = \mu_2^{2}, \quad m_{12}^{2} = 0,\quad
2\lambda_2 = 2\lambda_1= \lambda_3, \quad
\lambda_4 = \text{Re}(\lambda_5) = \lambda_6 = \lambda_7 = 0.
\label{eq:m-c}
\end{align}
The above potential is a functional of a single symmetric block,
$S_2$, i.e.~$V_{\text{MS-2HDM}} = V[S_2]$.

In addition, the potential term $V_{inert}$ in \eqref{sym+inert}
represents the inert scalar sector of the theory consists of $m$ inert
Higgs doublets $\widehat{\phi}_{\hat{i}}$ (with
$\hat{i}=\hat{1}, \hat{2}, \cdots, \hat{m}$), for which
$\langle \widehat{\phi}_{\hat{i}} \rangle=0$, and $N_H \equiv n-m$
Higgs doublets $\phi_{i}$ (with $i=1, 2, \cdots, N_H$) which generally
take part in EWSB with non-zero VEVs, $\langle\phi_{i} \rangle\neq0$.
The potential term $V_{inert}$ takes on the form
\begin{eqnarray}
	V_{\rm inert} &=& \widehat{m}_{\hat{i}\hat{j}}^{2} \widehat{\phi}_{\hat{i}}^{\dagger} \widehat{\phi}_{\hat{j}}
	              + \lambda_{\hat{i}\hat{j}\hat{k}\hat{l}} \left( \widehat{\phi}_{\hat{i}}^{\dagger} \widehat{\phi}_{\hat{j}} \right) \left( \widehat{\phi}_{\hat{k}}^{\dagger} \widehat{\phi}_{\hat{l}} \right)
	              + \lambda_{\hat{i}\hat{j}kl} \left( \widehat{\phi}_{\hat{i}}^{\dagger} \widehat{\phi}_{\hat{j}} \right) \left(\phi_{k}^{\dagger}\phi_{l} \right)
	\nonumber \\
	              &+& \lambda_{i\hat{j}\hat{k}l} \left(\phi_{i}^{\dagger} \widehat{\phi}_{\hat{j}} \right) \left( \widehat{\phi}_{\hat{k}}^{\dagger}\phi_{l} \right)
	              + \left[ \lambda_{i\hat{j}k\hat{l}} \left(\phi_{i}^{\dagger} \widehat{\phi}_{\hat{j}} \right) \left(\phi_{k}^{\dagger} \widehat{\phi}_{\hat{l}} \right) + {\rm H.c.}\right],
\end{eqnarray}
which must remain invariant under the $Z^I_{2}$ symmetry, \bea
Z^I_{2}: ~\phi_i~\to~\phi_i,
~\widehat{\phi}_{\hat{j}}~\to-\widehat{\phi}_{\hat{j}}.  \eea The
$m \times m$ matrix $\widehat{m}_{\hat{i}\hat{j}}^{2}$ is taken to be
positive definite, so as to avoid spontaneous EWSB.

Consequently, $Z^I_{2}$ should always be contained in the
$\mathcal{D}$ symmetry group of the inert scalar sector,
i.e.~$Z^I_{2}\subset \mathcal{D}$.  Notice that, in the case of Type-I
inert 2HDM, $Z^I_{2}$ can be extended by
another~$Z_2^{\text{EW}} \sim S_2$, namely the Permutation group which
consists of the field permutation:
$(\phi_1, \phi_2)\to (\phi_2, \phi_1)$. In general, the combined
action $Z_2^{\text{EW}}\times Z^I_{2}$ enforces SM alignment in the
$n$HDM, even beyond the tree-level
approximation~\cite{Pilaftsis:2016erj}.  Note that in the 2HDM the
constraint arising from $S_2\times Z^I_{2}$ symmetry meets
the alignment conditions given in Eq.\eqref{eq:NAcond2}, with
$\lambda_6=\lambda_7=0$.

Finally, the third term $\Delta V$ in \eqref{sym+inert} contains the
soft-symmetry breaking mass parameters of the EWSB sector and is given
by
\begin{align}
 \label{eq:Vsoft}
\Delta V\ =\ \displaystyle\sum_{i,j=1,2}^{N_H}\, m_{ij}^{2}\, ( \phi_i^{\dagger} \phi_j).
\end{align}
where $m_{ij}^{2}$ is in general a Hermitian $N_H \times N_H$ matrix,
with at least one negative eigenvalue in order to trigger EWSB. We assume that
the soft-symmetry breaking mass matrix $m_{ij}^{2}$ has no particular
discrete symmetry structure.

Therefore, the symmetries for SM alignment 
acting on the EWSM sector are~\cite{Pilaftsis:2016erj}
\begin{align}
&\text{(i) Sp(2}N_H) \times \mathcal{D}\;, \qquad
&\text{(ii) SU(2}N_H)  \times \mathcal{D},\qquad
&\text{(iii) SO(2}N_H) \times \mathcal{D},
\end{align}
where $N_H=n-m$ refers to the number of non-inert doublets and the
symmetry group~$\mathcal{D}$ has an effect only on the $m$ inert doublets. In
addition, as mentioned above, there is a minimal discrete
symmetry~\cite{Pilaftsis:2016erj}, 
\bea Z^{\text{EW}}_2\times Z^I_{2},  \eea
which can enforce exact SM alignment when satisfied by the complete Lagrangian.

In the next section, we will focus on the simplest realisation of SM alignment
i.e. the MS-2HDM.

\section{Maximally Symmetric 2HDM} \label{sec:3}

As we have seen in the previous sections, the SO$(5)$ symmetry puts
tight restrictions on the allowed form of the 2HDM potential, which
obeys naturally the experimental constraints that come from SM alignment.

After EW symmetry breaking, the following breaking pattern takes place:
\begin{align}
\text{SO(5)}\ \xrightarrow[]{\left< \Phi_{1,2} \right> \neq 0}\ \text{SO(4)}.
\end{align}
If $\Delta V$ vanishes, the CP-even scalar $H$ receives a non-zero mass
$M_H= v\sqrt{2 \lambda_2}$, while the other scalars, $h,\, a$ and
$h^\pm$, remain all massless with sizeable couplings to the SM gauge bosons.  These massless pseudo-Goldstone bosons~\cite{Goldstone}
 would open several experimentally excluded decay channels,
e.g.~$Z \to\,h\,a$ and $W^\pm \to h  h^\pm$~\cite{22}.  If the SO$(5)$
symmetry is realised at some high energy scale $\mu_X$ ($\gg
\mu_{\text{EW}}$), then due to RG running the following breaking
pattern will emerge~\cite{Dev:2014yca}:
\begin{eqnarray}
\text{SO(5)} \times \text{SU(2)}_L &\xrightarrow[]{g^{\prime}\neq 0}&
 \text{O(3)} \times \text{O(2)} \times \text{SU(2)}_L 
 \sim \text{O(3)} \times \text{U(1)}_Y \times \text{SU(2)}_L 
 \nonumber \\ &\xrightarrow[]{\text{Yukawa}}&
 \text{O(2)} \times \text{U(1)}_Y \times \text{SU(2)}_L
 \sim \text{U(1)}_{\text{PQ}} \times \text{U(1)}_Y \times \text{SU(2)}_L
 \nonumber \\ &\xrightarrow[]{\left< \Phi_{1,2} \right> \neq 0}&
 \text{U(1)}_{\text{em}}.
\end{eqnarray}
Note that the RG running of the gauge coupling $g^{\prime}$ only lifts
the charged Higgs mass $M_{h^\pm}$, while the corresponding effect of
the Yukawa couplings (particularly that of the top-quark~$h_t$)
renders the other CP-even pseudo-Goldstone boson~$h$ massive. Instead,
the CP-odd scalar $a$ remains massless and can be identified with a
Peccei$-$Quinn~(PQ) axion after the SSB of a global U$(1)_{\rm PQ}$
symmetry~\cite{Peccei:1977hh,Weinberg:1977ma,Wilczek:1977pj}.  Since
weak-scale PQ axions have been ruled out by experiment, we have
allowed for the SO$(5)$ symmetry of the MS-2HDM potential
in~\eqref{eq:VMS2HDM} to be broken by the soft SO(5)-breaking mass
term Re$(m^{2}_{12})$. With this minimal addition to the MS-2HDM potential, the
scalar-boson masses are given, to a very good approximation, by
\begin{align}
  \label{eq:m12}
 M_H^{2} = 2\lambda_2 v^{2}, \qquad M_h^{2}\ =\ M_a^{2}\ =\ M_{h^{\pm}}^{2}\ 
 =\ {\text{Re}(m_{12}^{2}) \over c_{\beta} s_{\beta} }.
\end{align}
Hence, all pseudo-Goldstone bosons, i.e. $h$, $a$ and $h^\pm$,  become
massive and almost degenerate in mass. 

In our study, the charged Higgs boson mass $M_{h^{\pm}}$ plays the
role of an input parameter in the ranges above $500$-GeV, in agreement
with $B$-meson constraints \cite{Misiak:2017bgg}. It will also be
considered as our threshold above which all parameters run with 2HDM
renormalisation group equations (RGEs). Note that, we implement the
matching conditions with two-loop RG effects of the SM at given
$M_{h^{\pm}}$ threshold scales.  Additionally, we employ two-loop 2HDM
RGEs to find the running of the gauge, Yukawa and quartic couplings at
RG scales larger than~$M_{h^{\pm}}$. For reviews on RGEs in the 2HDM,
see~\cite{Dev:2014yca,Oredsson:2018yho,Chowdhury:2015yja,Krauss:2018thf,Bednyakov:2018cmx}.
The SM and the 2HDM RGEs have been computed using the public
\texttt{Mathematica} package \texttt{SARAH} \cite{Sarah}, which has
been appropriately adapted for the MS-2HDM.

\subsection{Quartic Coupling Unification} \label{sec:3-1}

We have seen how the SO(5) symmetry of the MS-2HDM potential is broken
explicitly by RG running effects and soft-mass terms.  Here, we
consider a unified theoretical framework in which the SO(5) symmetry
is realised at some high-energy scale~$\mu_X$, where all the
conditions for the SM alignment are satisfied.  Of particular interest
is the potential existence of conformally-invariant unification points
at which all quartic couplings of the MS-2HDM potential vanish
simultaneously.

To address the above issue of quartic coupling unification, we employ
two-loop RGEs for the MS-2HDM from the unification scale $\mu_X$ to
the charged Higgs-boson mass $M_{h^{\pm}}$, where~$\mu_X \gg M_{h^{\pm}}$. Below this
threshold scale $\mu_{\rm thr} = M_{h^{\pm}}$, the SM is a viable
effective field theory, so we use the two-loop SM RGEs given
in~\cite{Carena:2015uoe} to match the relevant MS-2HDM couplings to
the corresponding SM quartic coupling $\lambda_{\rm SM}$, the Yukawa
couplings, the U(1)$_Y$ and SU(2)$_L$ gauge couplings $g_1$ and $g_2$
(with $g^{\prime} = \sqrt{3/5}g_1$). 

Note that the matching conditions for the Yukawa couplings at the threshold
scale read
\begin{equation}
 \label{eq:Ymatch}
 h_t^{\text{MS-2HDM}} = {y_t \over s_{\beta}}, \qquad
h_b^{\text{MS-2HDM}} = {y_b \over c_{\beta}}, \qquad
h_{\tau}^{\text{MS-2HDM}} = {y_{\tau} \over c_{\beta}}.
\end{equation}
For higher RG scales $\mu > \mu_{\rm thr}$, the
running of the Yukawa couplings $h_t$, $h_b$ and $h_\tau$ is
governed by two-loop 2HDM RGEs.

\begin{figure}[!t]
\centering
\includegraphics[width=0.8\textwidth]{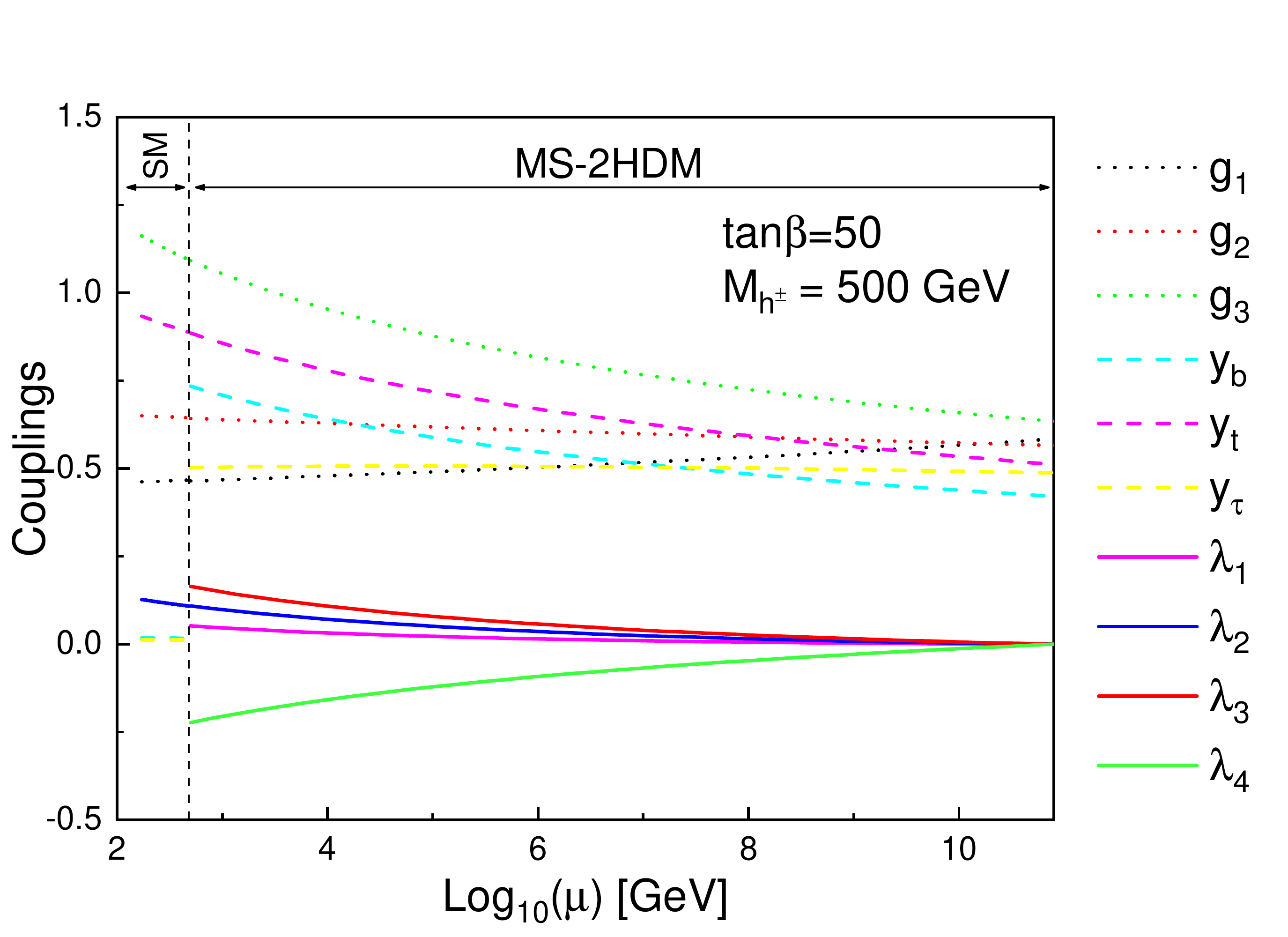}
\caption{\it The RG running of the quartic couplings
  $\lambda_{1,2,3,4}$, gauge and Yukawa couplings from the threshold
  scale $M_{h^\pm} = 500$ GeV up to the  their first quartic coupling unification
  scale~${\mu^{(1)}_X=10^{11}}$\,GeV for $\tan\beta = 50$.}
\label{0.5TeV}
\end{figure}
\begin{figure}[!h] 
\centering
\includegraphics[width=0.8\textwidth]{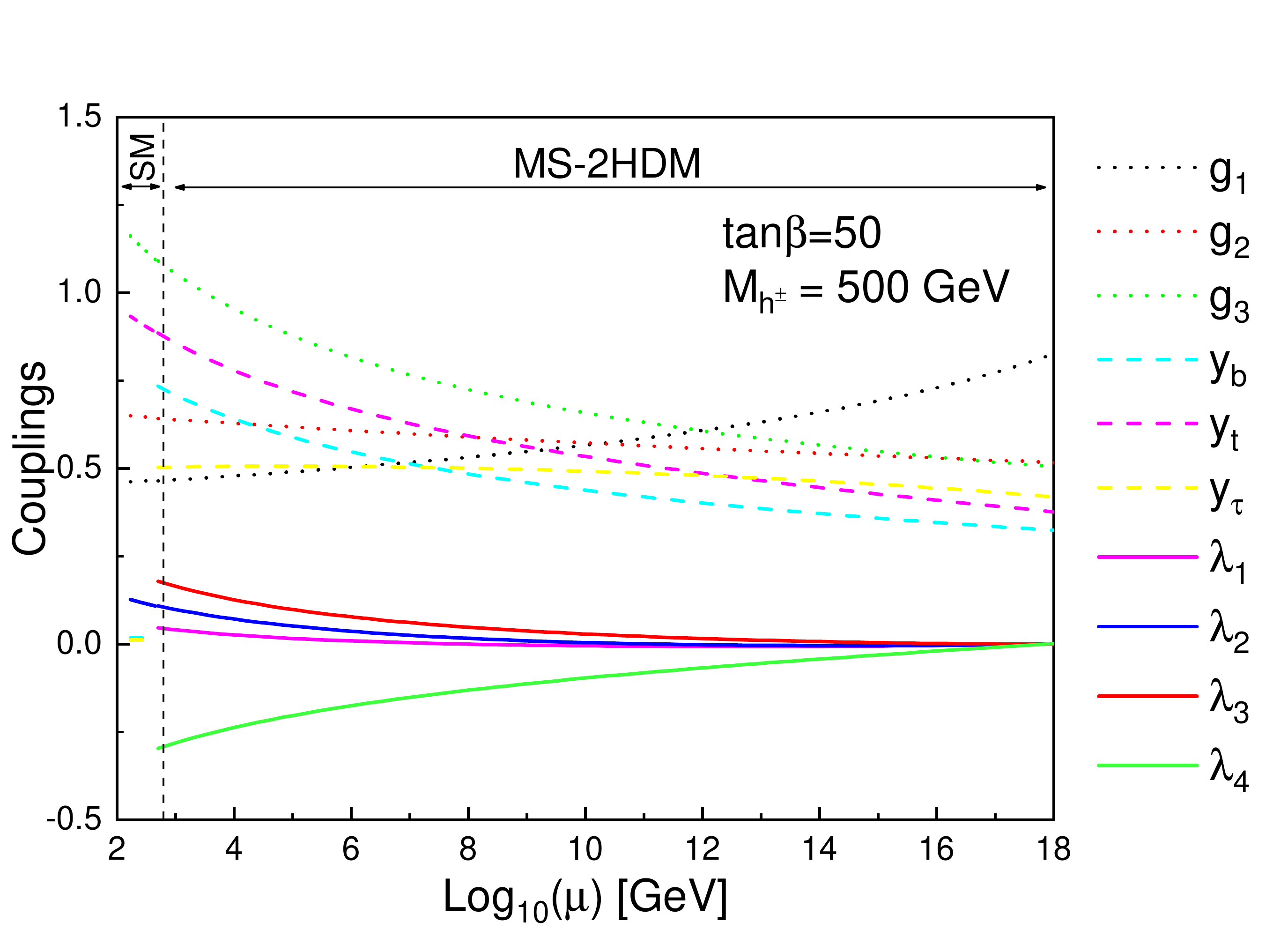}
\caption{\it The RG evolution
  is extended up to the second quartic coupling unification point
  $\mu^{(2)}_X=10^{18}$\,GeV.}
\label{0.5TeV2}
\end{figure}

Figures~\ref{0.5TeV}~and~\ref{0.5TeV2} exhibit the RG evolution of all
relevant couplings of the SM and the $\text{MS-2HDM}$,
for~$\tan\beta=50$ and $M_{h^\pm} = 500$~GeV. The vertical dashed line
shows the threshold scale $\mu_{\rm thr} = M_{h^\pm} = 500$~GeV.
Visible are at this scale significant discontinuities in the RG
running of Yukawa couplings $h_b$ and $h_\tau$ due to the matching
conditions.

In Figure~\ref{0.5TeV}, the quartic coupling $\lambda_2$, which
is correlated to the mass of SM-like Higgs-boson~$M_H$, decreases at high RG
scales due to the evolution of the top-Yukawa coupling $h_t$ and turns
negative just above the quartic coupling unification scale
$\mu_X \sim 10^{11}$~GeV, where all quartic couplings vanish.
Thereby, for energy scales above the RG scale $\mu_X$, we envisage
that the MS-2HDM will need to be embedded into another UV-complete
theory. Nevertheless, according to our estimates
in~\cite{Darvishi:2019ltl}, we have checked that the resulting MS-2HDM
potential leads to a metastable but sufficiently long-lived EW vacuum,
whose lifetime is many orders of magnitude larger than the age of our
Universe. In this respect, we regard the usual constraints derived
from convexity conditions on 2HDM potentials~\cite{Deshpande:1977rw}
to be over-restrictive and unnecessary for our theoretical framework.

Of equal importance is a second conformally-invariant unification
point at energy scales close to the reduced Planck mass
$\mu^{(2)}_X\sim10^{18}$\,\text{GeV}, as shown in Figure~\ref{0.5TeV2}. In
this case, the key quartic coupling~$\lambda _2$ increases and turns
positive again. Therefore, in this class of settings, any embedding of
the MS-2HDM into a UV-complete theory must have to take quantum
gravity into account as well.

\begin{figure}[t]
\centering
\includegraphics[width=0.87\textwidth]{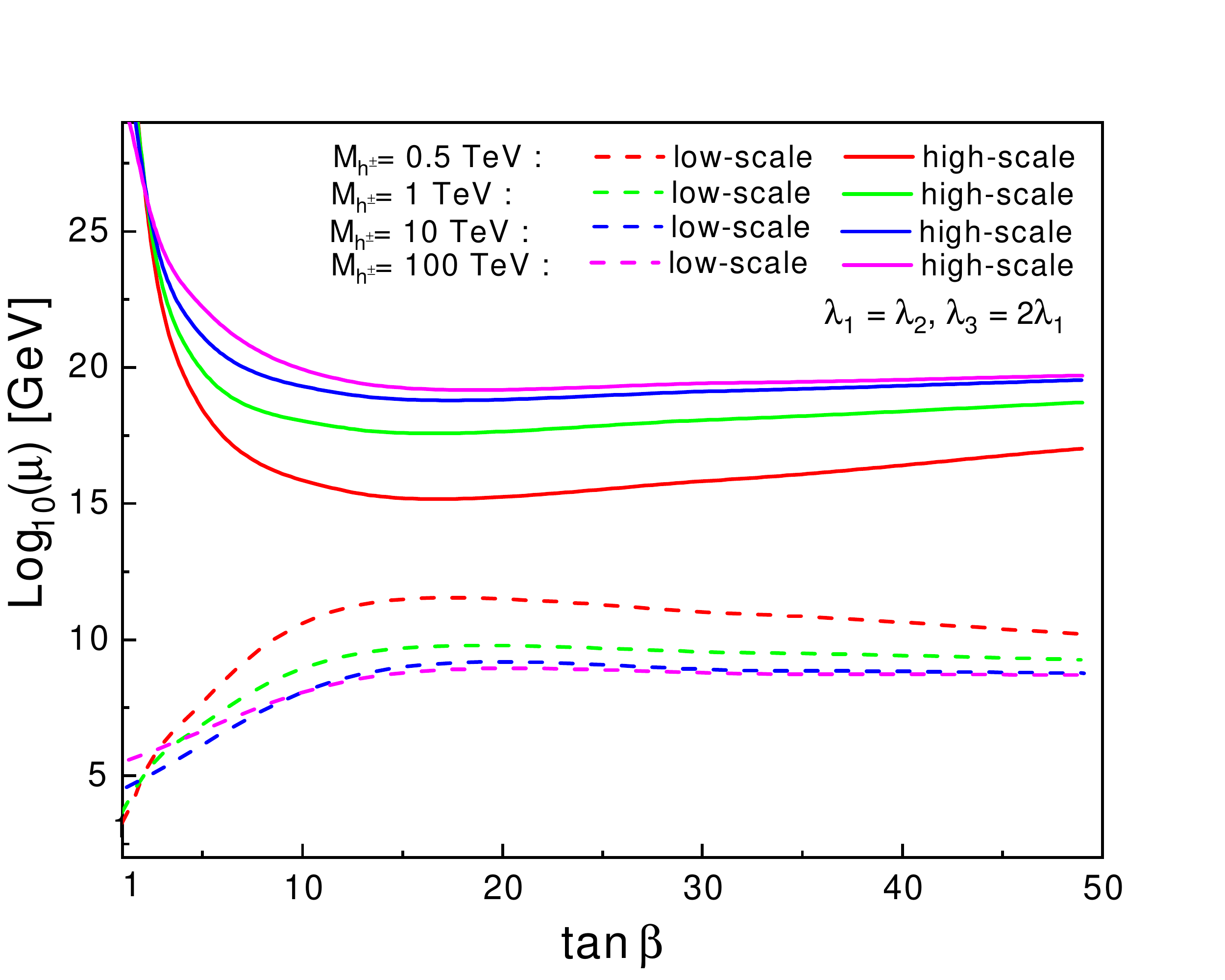}
\caption{\it Sets of quartic coupling unification points in the
  $(\tan\beta,\, \log_{10}\mu )$ plane, for various values of charged Higgs-boson masses
  $\,M_{h^{\pm}}=500\,\textit{GeV},\,1\,\textit{TeV},\,10\,\textit{TeV}\,
\textit{and}\,100\,\textit{TeV}$.  The dashed and solid curves show the sets of low-scale 
$\mu^{(1)}_X$ and high-scale $\mu^{(2)}_X$ quartic coupling unification points.}
\label{mix2}
\end{figure}

By analogy, Figure~\ref{mix2} shows all conformally-invariant quartic coupling
unification points in the $(\tan\beta,\,\log_{10}\mu)$ plane, by
taking into account different values of threshold scales~$\mu_{\rm thr}$,
i.e.~for $\mu_{\rm thr}\, =\,M_{h^{\pm}}=500\,\text{GeV},\,1\,\text{TeV},\,10\,\text{TeV}\,
\text{and}\,100\,\text{TeV}$. The lower curves (dashed curves) correspond to sets
of low-scale quartic coupling unification points, while the upper
curves (solid curves) give the corresponding sets of high-scale
unification points. From Figure~\ref{mix2}, we may also observe the
domains in which the $\lambda_2$ coupling becomes negative. These
are given by the vertical $\mu$-intervals bounded by the lower and the
upper curves, for a given choice of $M_{h^\pm}$ and
$\tan\beta$. Evidently, as the threshold scale~$\mu_{\rm thr}=M_{h^\pm}$
increases, the size of the negative $\lambda_2$ domain increases and
becomes more pronounced for smaller values~of~$\tan\beta$.

Having gained valuable insight from this model, it is important to highlight that the
MS-2HDM requires only three additional input parameters: (i)~the soft
SO$(5)$-breaking mass parameter $m^{2}_{12}$ (or $M_{h^{\pm}})$, (ii)~the
ratio of VEVs $\tan\beta$, and (iii) the conformally-invariant quartic
coupling unification scale~$\mu_X$ which can only assume two discrete
values: $\mu^{(1)}_X$ and $\mu^{(2)}_X$. Knowing these 
parameters, the entire Higgs sector of the model can be
determined. In the next part, we will give typical
predictions in terms of these three input parameters.

\subsection{Misalignment Predictions for Higgs Boson Couplings}\label{misa} 

As was discussed in Section \ref{sec:1}, the {\em misalignment} of the SM-like Higgs-boson couplings $HVV$ (with
$V= W^\pm,\,Z$), $Ht\bar{t}\, \text{and}\,Hb\bar{b}$
are controlled by the light-to-heavy scalar-mixing parameter
$\theta_{\mathcal{S}}$. 
Obviously, at the quartic coupling unification scale
$\mu_X$, the SO$(5)$ symmetry of the MS-2HDM is fully restored and this
mixing parameter vanishes. However, 
RG effects induced by the U(1)$_Y$ gauge coupling and the Yukawa
couplings break sizeably the SO$(5)$
symmetry, giving rise to a calculable \text{non-zero} value for
$\theta_\mathcal{S}$ and thereby to misalignment predictions for {\em
 all} $H$-boson couplings to \text{SM\,particles}. Here, we present numerical estimates of the predicted
deviations of the SM-like Higgs-boson couplings $HVV$ (with
$V= W^\pm,\,Z$), $Ht\bar{t}\, \text{and}\,Hb\bar{b}$, from their respective SM
values.

\begin{figure}[t]
\centering
\label{1stc}
\includegraphics[width=0.49\textwidth]{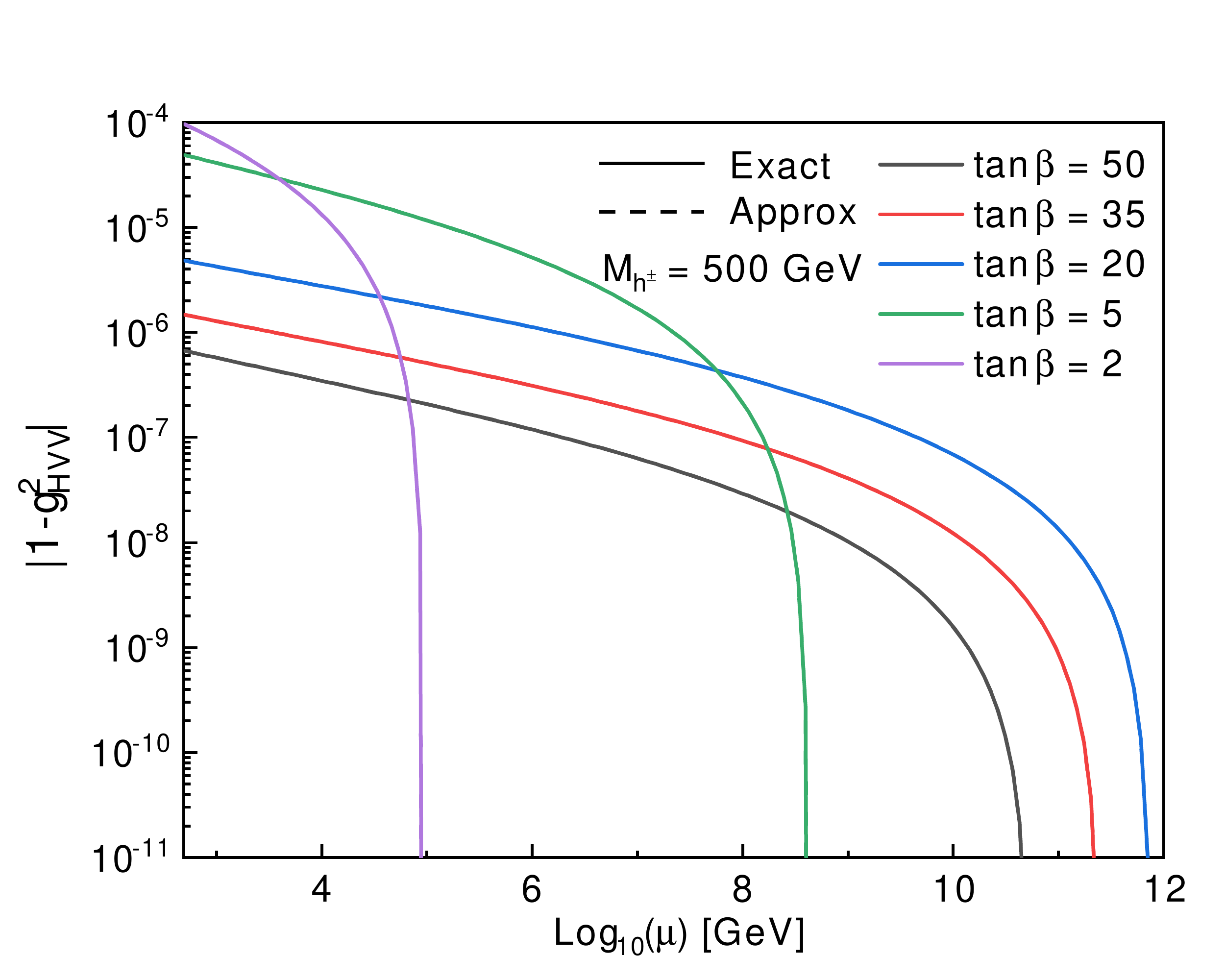}
\includegraphics[width=0.49\textwidth]{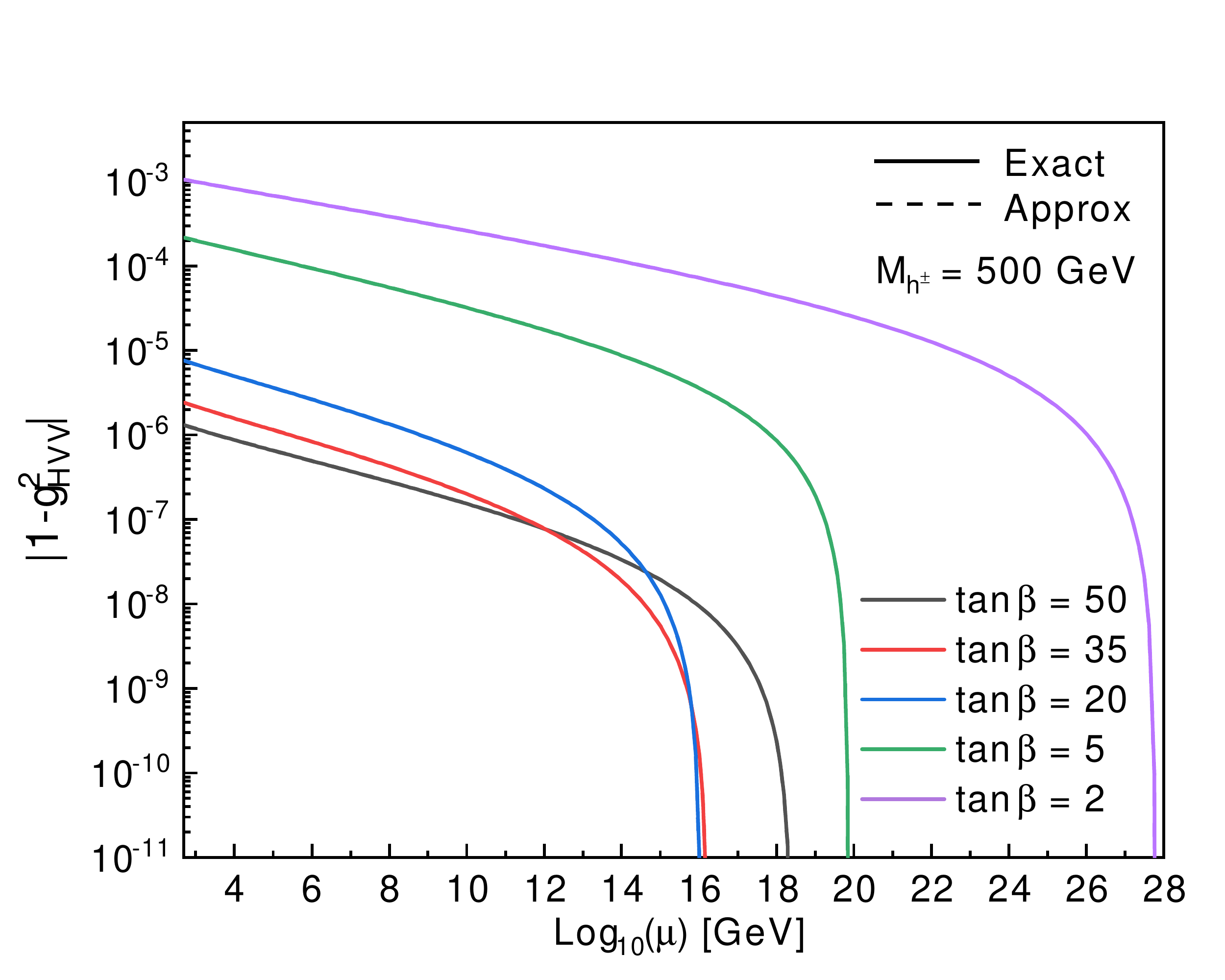}
\caption{\it Numerical estimates of the misalignment parameter $|1 -
  g^{2}_{HVV}|$ pertinent to the $HVV$-coupling (with $V= W^\pm,Z$) as
  functions of the RG scale~$\mu$, for low-scale (left panel) and high-scale (right panel) quartic coupling
  unification scenarios, assuming  $M_{h^\pm} = 500$\,GeV and
  $\tan\beta=2,\ 5,\,20,\,35$ and $50$.}
\end{figure}

The dependence of the physical misalignment parameter
\text{$|1 - g^{2}_{HVV}|$} (with $g_{H_{\text{SM}}V V} =1$) as functions
of the RG scale~$\mu$ is shown in Figure~\ref{1stc}.  This is given
for typical values of $\tan\beta$, such as
$\tan\beta=2,\ 5,\,20,\,35\,\text{and}\,50$.  As expected, the
normalised coupling $g_{HVV}$ approaches the SM value
$g_{H_{\text{SM}}V V} =1$ at the lower- and higher-scale quartic
coupling unification points, $\mu^{(1)}_X$ and $\mu^{(2)}_X$, as shown
in left and right panels, respectively.  We use dashed lines to
display our predictions to leading order in $\theta_\mathcal{S}$
expansion, while solid lines stand for the exact all-orders
result. Since there is a small deviation (below the per-mile level) of
$g_{HVV}$ from the SM value, the approximate and exact predicted
values are predominately overlapping.  Evidently, the misalignment
reaches its maximum value for low values of $\tan \beta$ and for the
higher quartic coupling unification points.

\begin{figure}[h]
\centering
\includegraphics[width=0.49\textwidth]{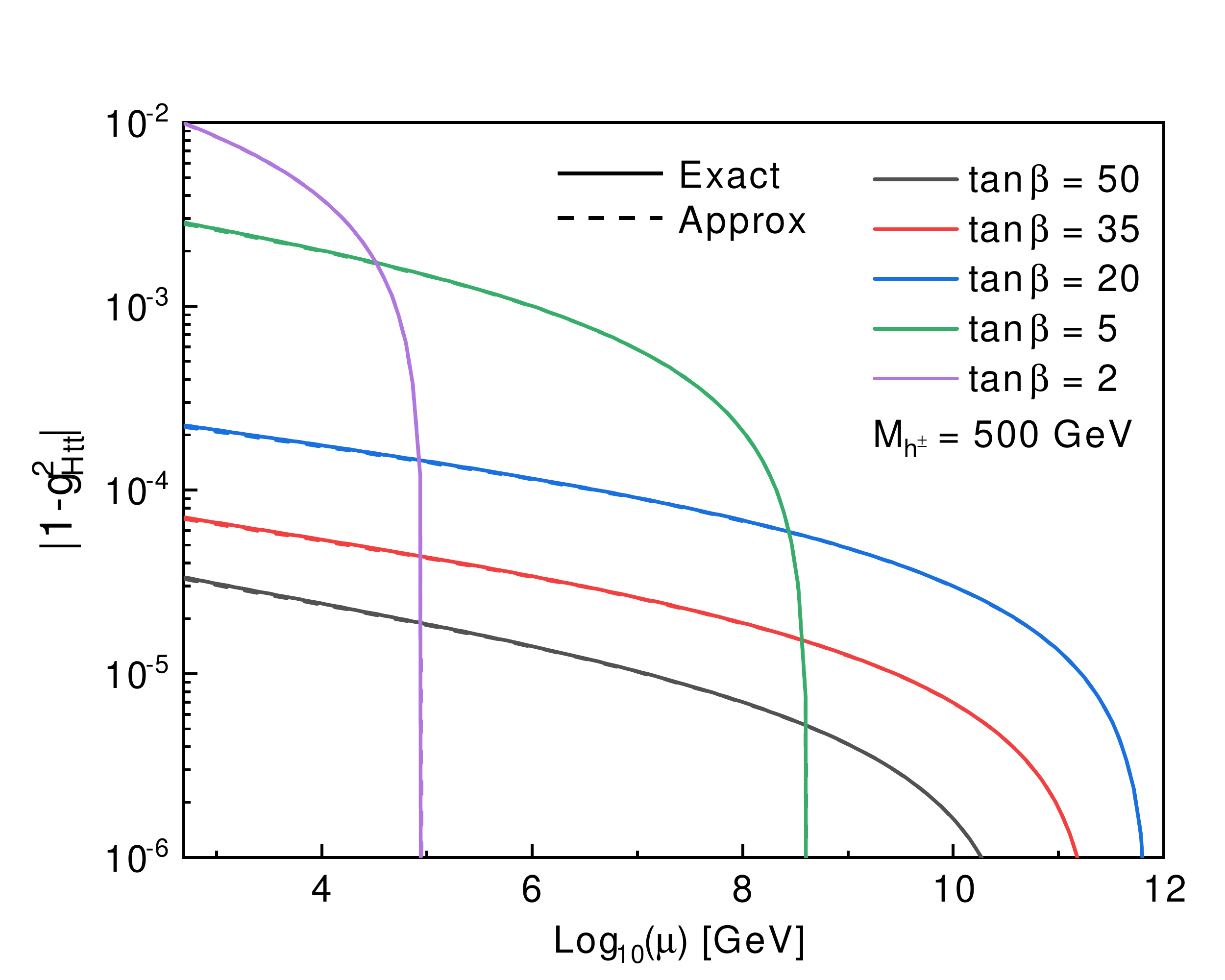}
\includegraphics[width=0.49\textwidth]{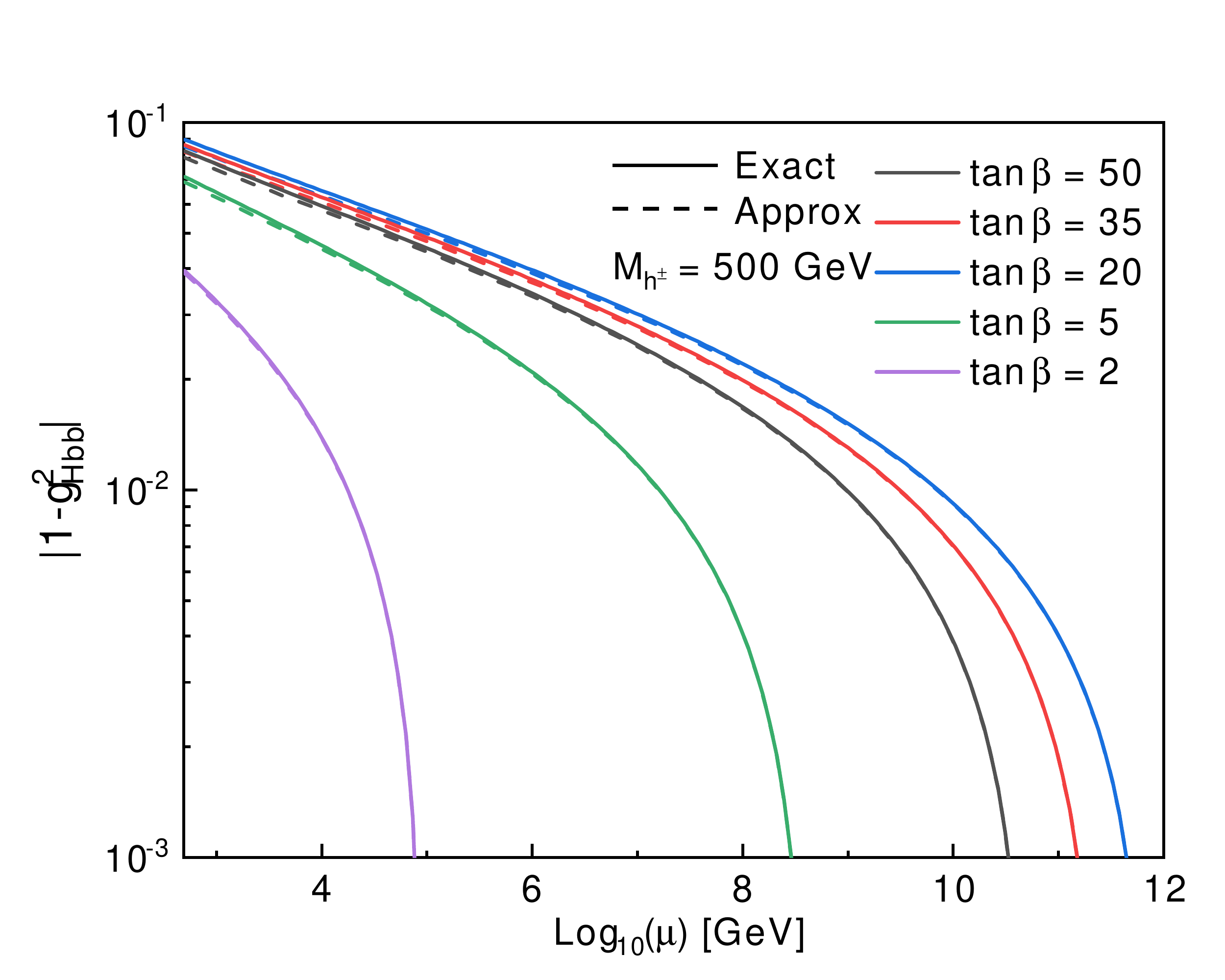}
\caption{\it Numerical estimates of the misalignment parameters $|1 -
  g^{2}_{Htt}|$ (left panel) and $|1 -
  g^{2}_{Hbb}|$ (right panel) versus the RG scale~$\mu$, for a  low-scale quartic coupling
  unification scenario, assuming  $M_{h^\pm} = 500$\,GeV and
  $\tan\beta=2,\ 5,\,20,\,35$ and $50$.}
\label{1stcf}
\centering
\end{figure}

\begin{figure}[t]
\includegraphics[width=0.49\textwidth]{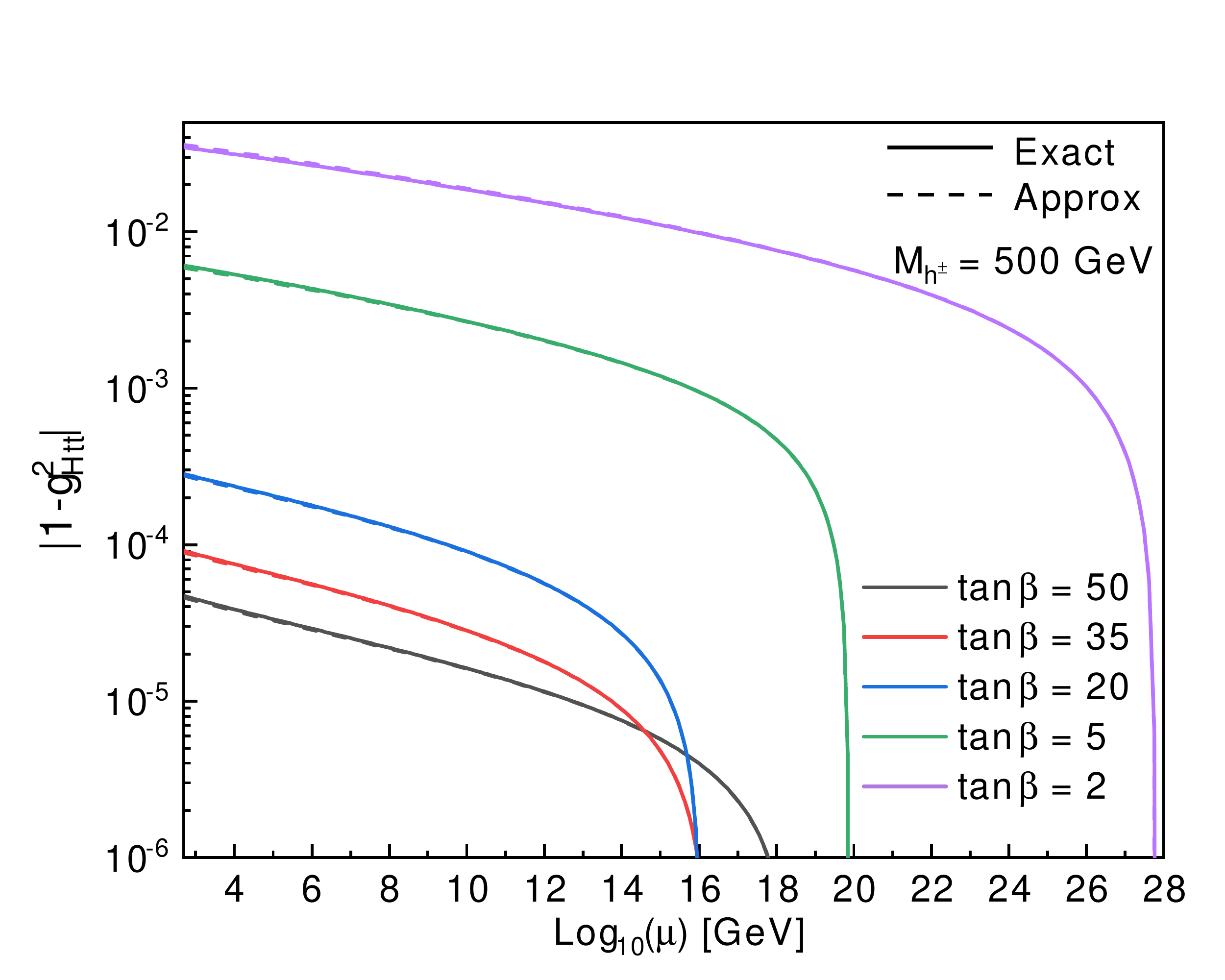}
\includegraphics[width=0.49\textwidth]{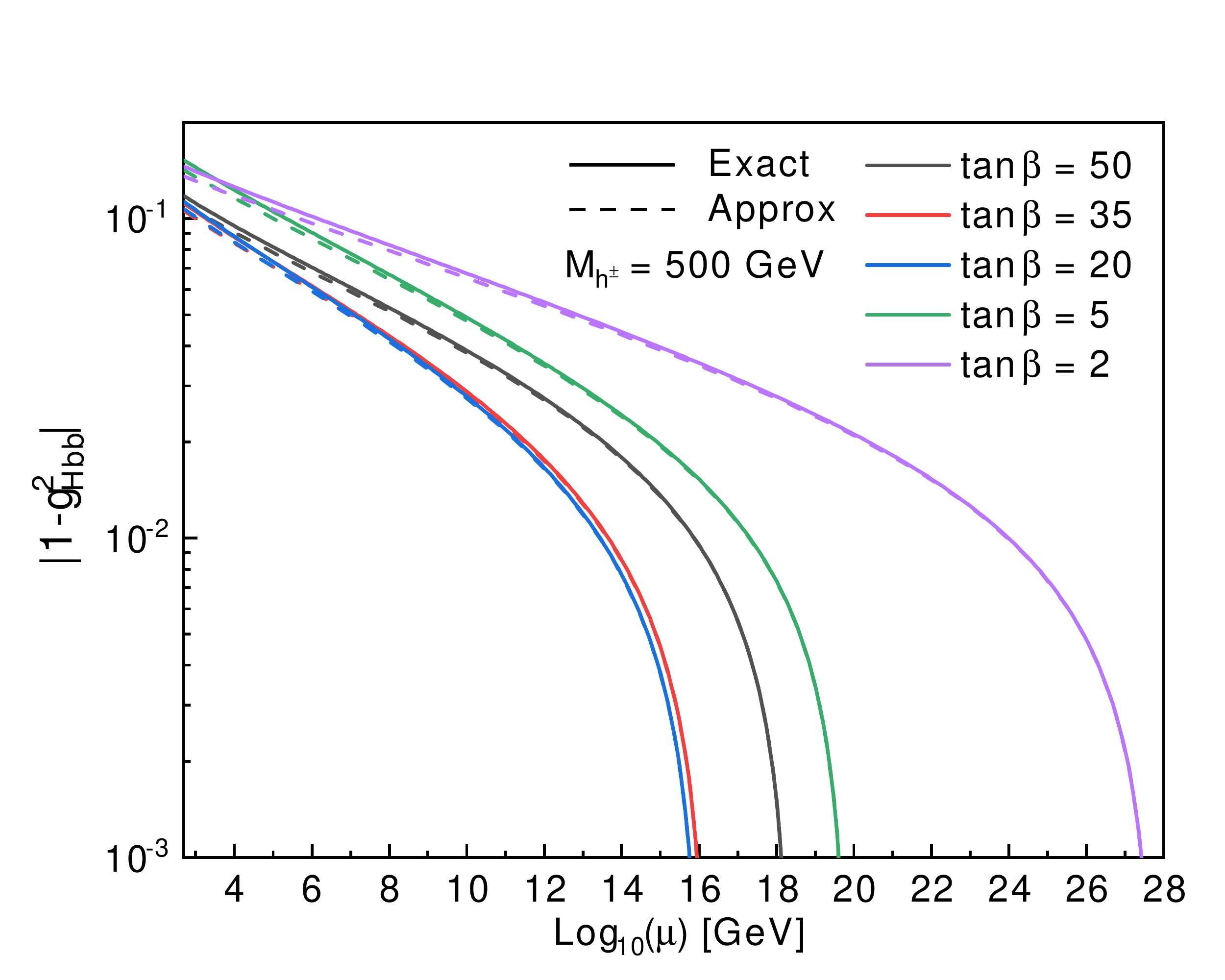}
\caption{\it The same as in Figure~5, but for a high-scale quartic coupling
  unification scenario.} 
\label{2ndcf}
\end{figure}

By analogy, Figures \ref{1stcf} and \ref{2ndcf} display misalignment
predictions for the $H$-boson couplings to top- and bottom-quarks, for
$\tan\beta=2,\ 5,\,20,\,35$ and $50$ and for lower- and higher-scale
quartic coupling unification points, respectively.  As before, the
deviation of the normalised couplings $g_{Htt}$ and $g_{Hbb}$ from
their SM values are larger for low values of $\tan \beta$,
e.g.~$\tan\beta = 2$, and for higher-scale quartic coupling unification
points~$\mu^{(2)}_X$. This effect is more noticeable for~$g_{Hbb}$, as the amount of
misalignment might be even larger than $10\%$. In this case, a comparison
between solid and dashed lines shows the appropriateness of our
seesaw-inspired approximation in terms of~$\theta_\mathcal{S}$ parameter.
 
Last but not least, we confront our misalignment predictions for the SM-like
Higgs boson couplings, $g_{HZZ}$, $g_{Htt}$ and $g_{ Hbb}$ with
existing experimental data from ATLAS and CMS, including their
statistical and systematic uncertainties~\cite{201606}. Our
predictions for $\tan\beta=2,\,20,\,35\,\text{and}\,50$ and $M_{h^\pm}=500$~\text{GeV} are presented in Table \ref{ex}.
The observed results for $g_{HZZ}$ and $g_{Htt}$ are in excellent consistency with the SM and the MS-2HDM. Instead, the LHC data for
$g_{Hbb}$ can be fitted to the SM at the $3\sigma$ uncertainty level. Interestingly, the uncertainty level
reduces only to $2\sigma$ in the case of MS-2HDM, for $\tan\beta=2$
assuming a high-scale quartic coupling unification scenario.  Future 
precision collider experiments might be able to probe such a scenario.


\begin{table*}[t]
 \centering
 \small
 \begin{tabular}{c c c c c c c c c c c c c c c c c c c c c c c c c c }
 \hline
 Couplings&&ATLAS&&CMS&&$t_\beta = 2$&&$t_\beta = 5$&&$t_\beta = 20$&&$t_\beta = 50$\\ \hline
 $|g_{HZZ}^{\mu_X^{(1)}}|$&&[0.86, 1.00]&& [0.90, 1.00]&&0.999&&0.999&&0.999&&0999 \\
 $|g_{HZZ}^{\mu_X^{(2)}}|$ && && && 0.998&&0.999&&0.999 &&0.999\\ \hline
 $|g_{Htt}^{\mu_X^{(1)}}|$&&$1.31^{+ 0.35}_{- 0.33}$&&$1.45^{+0.42}_{-0.32}$&&1.004&&1.001&&1.000&&1.000 \\
 $|g_{Htt}^{\mu_X^{(2)}}|$&& && &&1.098&&1.017&&1.000 &&1.000 \\ \hline
 $|g_{Hbb}^{\mu_X^{(1)}}|$&&$0.49^{+ 0.26}_{- 0.19}$&&$0.57^{+0.16}_{-0.16}$&&0.980&& 0.964&&0.956 &&0.959\\
 $|g_{Hbb}^{\mu_X^{(2)}}|$&& && && 0.881&&0.926&&0.944 &&0.942 \\
 \hline
 \end{tabular}
 \caption{\it Predicted values of the SM-like Higgs boson  couplings
 to the $Z$ boson and to top- and bottom-quarks in the MS-2HDM for both
 scenarios with low- and high-scale quartic coupling unification,
 assuming $M_{h^{\pm}}=500$\,GeV. The corresponding central values
 for these couplings from ATLAS and CMS are also given, including their
 uncertainties~\cite{201606}. } 
 \label{ex}
\end{table*}

\newpage

\section{Conclusions} \label{con}

The potential of $n$-Higgs Doublet Models ($n$HDMs), with $n \geq 2$
Higgs doublets contains a large number of SU(2)$_L$-preserving
accidental symmetries as subgroups of the symplectic group Sp(2$n)$.
We have identified \textit{the complete set} of maximal continuous
symmetries for the so-called SM alignment that may take place in the
$n$HDM potentials.  These symmetries are necessary for
\textit{natural} SM alignment, without decoupling of large mass scales
or fine-tuning. For instance, for the 2HDM, these symmetries ensure
alignment for {\em any} value of $\tan\beta$ and {\em any} form of
soft-breaking by bilinear mass terms of the scalar potential.  In
general, the Maximal Symmetric $n$HDM (MS-$n$HDM) can provide
natural SM alignment exhibiting quartic coupling unification up to the
Planck~scale.

For our illustrations, we have analyzed the simplest realisations of a
Type-II 2HDM, the so-called Maximally Symmetric Two-Higgs Doublet
Model (MS-2HDM).  The scalar potential of this model is determined by
 a single mass parameter and a single quartic coupling. This minimal
form of the potential can be reinforced by an accidental SO$(5)$
symmetry in the bilinear field space, which is isomorphic to Sp(4) in
the field basis $\bm{\Phi}$ given in~\eqref{eq:bfPhi}. The MS-2HDM can
naturally realise the SM alignment limit, in which all SM-like Higgs
boson couplings to the EW gauge bosons and to all fermions are
equal to their SM strength independently of the values of $\tan\beta$ and $M_{h^{\pm}}$.

The SO$(5)$ symmetry of the MS-2HDM is broken explicitly by RGE effects
due to the non-zero U(1)$_Y$ gauge coupling and equally sizeably by the Yukawa
couplings.  For
phenomenological reasons, we have also added a soft SO$(5)$-breaking
mass parameter~$m^{2}_{12}$ to the scalar potential, which
lifts the masses of all pseudo-Goldstone bosons $h^\pm$, $h$ and $a$
in the ranges above 500-GeV, consistent with $B$-meson constraints.  

To evaluate the RG running of the quartic couplings and the relevant SM
couplings, we have employed two-loop 2HDM RGEs from the unification
scale $\mu_X$ up to charged Higgs-boson mass~$M_{h^\pm}$. At the RG
scale~$M_{h^\pm}$, we have implemented matching conditions between the
MS-2HDM and the SM parameters. 

Improving upon an earlier study~\cite{Dev:2014yca}, we have now
clearly demonstrated that in the MS-2HDM all quartic couplings can
unify at much larger RG scales $\mu_X$, where $\mu_X$ lies between
$\mu^{(1)}_X \sim 10^{11}\,$~GeV and $\mu^{(2)}_X \sim 10^{20}\,$~GeV.
In particular, we have shown that quartic coupling unification can
take place in two different conformally invariant points, at which all
quartic couplings vanish.  This property is unique for this model and
can happen at different threshold scales
$\mu_{\rm thr}\,
=\,M_{h^{\pm}}=500\,\text{GeV},\,1\,\text{TeV},\,10\,\text{TeV}\,
\text{and}\,100\,\text{TeV}$.
More precisely, the low-scale (high-scale) unification point arises
when $\lambda_2$ crosses zero and becomes negative (positive) at the
RG scale $\mu_X^{(1)}$ ($\mu_X^{(2)}$).  Between these two RG scales,
i.e.~for $\mu^{(1)}_X < \mu < \mu^{(2)}_X$, the running quartic
coupling $\lambda_2(\mu )$ turns negative, which gives rise to a
deeper minimum in the effective MS-2HDM potential, whose lifetime is
extremely large much larger than the age of our
Universe~\cite{Darvishi:2019ltl}.

In summary, the MS-2HDM is a very predictive extension of the SM, as
it is governed by two additional parameters: (i)~the charged
Higgs-boson mass~$M_{h^{\pm}}$ (or $m^{2}_{12}$) and (ii)~the ratio of
VEVs~$\tan\beta$.  Moreover, requiring conformally-invariant quartic
coupling unification at scale~$\mu_X$, we find to two discrete values
for $\mu_X$, $\mu^{(1)}_X$ and $\mu^{(2)}_X$, as functions of
$M_{h^{\pm}}$ and $\tan\beta$. Once the values of these theoretical
parameters are known, the entire Higgs sector of the model can be
determined.  In this context, upon the fact that the RG effects
sizeably break the SO$(5)$ symmetry, we obtained a
calculable~\text{non-zero} value for the light-to-heavy scalar-mixing
parameter $\theta_{\mathcal{S}}$. Thereby, we have been able to
present illustrative predictions of misalignment for the SM-like
Higgs-boson couplings to the $W^\pm$ and $Z$ bosons and to the top-
and bottom-quarks.  The predicted deviations to $Hb\bar{b}$-coupling
is of order $10\%$ and may be observable at future $e^+e^-$ colliders.

\section*{Acknowledgements}

\noindent
The work of AP and ND is supported in part by the
Lancaster–Manchester–Sheffield Consortium\- for Fundamental Physics,
under STFC research grant ST/P000800/1.

\vfill\eject

\end{document}